\pdfoutput=1
\def\preprint{0}                
\def\preprint{1}                
\def\comment#1{}
\if\preprint1
        \documentclass[useAMS,usenatbib]{mn2e}
        \usepackage{times}
        \usepackage{graphicx}
        \usepackage{calc}
        \usepackage{multirow}
        \usepackage{multicol}
        \usepackage[flushleft]{threeparttable}
		\usepackage[singlelinecheck=false]{caption}
		\usepackage{amsmath}
		\usepackage{listings}
		\usepackage{natbib}	
		\usepackage{url}

\else
        \documentstyle[astrop-bib,referee,times]{mn}
        \newcommand{\includegraphics}[1]{}
\fi

\def\oversim#1#2{\lower0.5pt\vbox{\baselineskip0pt \lineskip-0.5pt
     \ialign{$\mathsurround0pt #1\hfil##\hfil$\crcr#2\crcr\sim\crcr}}}

{\title[Flickering in AGB stars]
	{Flickering in AGB stars: probing the nature of accreting companions}
\author[S. Snaid  et al.]{
S. Snaid$^{1,2}$,
A~A.~Zijlstra$^1$\thanks{E-mail: \tt a.zijlstra@manchester.ac.uk}, 
I. McDonald$^1$, 
Helen Barker$^1$,               
T. R. Marsh$^3$, 
V. S. Dhillon$^{4,5}$
\\
        $^1$Jodrell Bank Centre for Astrophysics, 
          School of Physics \&\ Astronomy, University of Manchester
          Oxford Road, Manchester M13 9PL, UK\\
        $^2$Astronomy Department, Faculty of Science, King Abdulaziz University, Jeddah 21589, Saudi Arabia\\
        $^3$Department of Physics, University of Warwick, Gibbet Hill Road, Coventry, CV4 7AL, UK\\
        $^4$Department of Physics and Astronomy, University of Sheffield,
        Sheffield S3 7RH, UK\\
        $^5$Instituto de Astrofisica de Canarias, E-38205 La Laguna, Tenerife, Spain        
}

\pubyear{2018}

\begin{document}

\maketitle
\begin{abstract}
Binary companions to asymptotic giant branch (AGB) stars are an important aspect of their evolution. Few AGB companions have been detected, and in most cases it is difficult to distinguish between main-sequence and white dwarf companions. Detection of photometric flickering, a tracer of compact accretion disks around white dwarfs, can help identify the nature of these companions. In this work, we searched for flickering in four AGB stars suggested to have likely accreting companions. We found no signs for flickering in two targets: R~Aqr and V1016 Cyg. Flickering was detected in the other two stars: Mira and Y Gem. We investigated the true nature of Mira's companion using three different approaches. Our results for Mira strongly suggest that its companion is a white dwarf.
\end{abstract}

\begin{keywords}
 Stars:  Binary systems, AGB, light curves -- Accretion disks: flickering -- Observations: ULTRACAM
 
\end{keywords}

\begin{table*}
\caption{Log of observations}
\label{Observations}
\begin{tabular}{ccccccc}
\hline
\hline
\multirow{2}{*}{Target}   & RA  & DEC  &  Date   & Start of run &    Obs. time   & 	\multirow{2}{*}{Airmass}  \\
& (hh:mm:ss) & (dd:mm:ss) & (Start of run) & (UT)  & (hours)  &  \\
\hline
V1016 Cyg & 19:57:05.0 & +39:49:36.3 & 27/09/2015  & 20:21:43 & 2.71  & 1.02-1.16  \\
R Aqr  & 23:43:49.5 & -15:17:04.1   & 27/09/2015   & 23:22:47   & 1.59   & 1.39-1.46   \\
Mira  & 02:19:20.8  & -02:58:39.5   & 28/09/2015  & 01:17:29  & 3.49  & 1.18-1.32  \\
&   &    & 29/09/2015   & 02:58:01   & 2.03   & 1.18-1.36   \\
Y Gem  & 07:41:08.5  & +20:25:44.3  & 29/09/2015   & 05:09:59    & 1.19   & 1.14-1.40     \\
\hline
\hline 
\end{tabular}
\end{table*}

\section{Introduction}
\indent Low and intermediate mass stars, 1 to 8\,M$_\odot$, end their lives by mass loss. During their post main-sequence (MS) evolution, these stars experience a phase of catastrophic stellar winds, when up to 80\%\ of the mass can be ejected. The mass loss occurs on the AGB phase, and is driven in part by Mira pulsations. The remnant core of the star evolves to become a white dwarf (WD), while the ejecta become ionized and form a planetary nebula (PN).\\
\indent Eighty percent of planetary nebulae (PNe) show asphericity in their morphologies \citep{MASH}. The precise mechanism behind this is still disputed. It has been suggested that the presence of a companion star can shape the mass loss from the PN or from the progenitor AGB star (e.g. \citealt{Nordhaus&Blackman06}, \citealt{Garcia-Segura+14}, \citealt{Zijlstra15}). However, the effect of binary companions on shaping winds depends on the binary separation. For the most compact systems ($ < $10AU), a common envelope can form, leading to extreme morphologies, but this is relatively rare \citep{Zijlstra2007}.
On the other hand, very wide binaries affect the shape only through orbital motion, and resulting spiral waves \citep{Nordhaus2010}. In between, at separations of 10-100AU, the dominant model is that of \cite{SahaiTrauger1998} who argue that the companions develop accretion disks and jets, and that these jets shape the outflows. Numerical models confirm that up to 10\%\ of the ejecta may be captured and form a disk \citep{Huarte-Espinosa2013}. The role of jets is more controversial, but jet-like flows have been seen in some binary AGB stars \citep[e.g. R~Aqr, ][]{Dougherty1995}. However, direct observations of companions are mostly lacking because AGB stars are very luminous ($\sim$$10^{3}$-$10^{4}\rm L_\odot$) and far outshine any companions (which are generally less luminous MS stars or WDs).\\
\indent A few binary companions are known, primarily in symbiotic stars where a hot companion affects the stellar wind. The best case study is the Mira system (Mira AB), which includes a companion (Mira~B) and shows high-velocity lobes, possibly arising from the companion \citep{Meaburn2009}. \cite{Sokoloski2010} argue for an accreting WD, based on observations of flickering, while \cite{Ireland2007} confirmed a 10AU accretion disk but argue for a K5 dwarf. Recently, \cite{Vlemmings2015} were able to resolve the partially ionized region around Mira~B at millimetre wavelengths and their observations confirmed that the 2.4AU region results from accreted material from the AGB wind.
Knowing the nature of the companion is important to the shaping mechanism. Both the accretion luminosities and the jet velocities depend on the depth of the gravitational potential. For an MS star, the effect of accretion is likely limited, and the dominant shaping mechanism may be orbital motion of the Mira. For a WD companion, significant energetic outflows may be expected. Clearly, our understanding is limited if we can't distinguish WD from MS companions.\\
\indent Compact accretion disks around WDs give rise to rapid fluctuations. Accretion disks give variable emission with a timescale that depends on the radius of the system and the mass of the accretor. Emission of the accretion disk is overwhelmed by the much brighter AGB star. The AGB star has a red spectral energy distribution (SED) while the accretion disk is blue. The accretion disk is, therefore, easier to detect in the blue. For Mira variables, observations during the minimum of the the light curve will further optimise the sensitivity to the accretion disk. Based on this, \cite{Warner72} and \cite{Sokoloski2010} detected variability in Mira, at timescales of minutes and longer. . However, the variability was not confirmed by other observations  \citep{Prieur2002} at timescales of 5-10 minutes.\\
\indent We have carried out observations of four AGB stars known to have binary companions, where the companion is likely accreting (symbiotic Miras). The four systems are Mira, Y Gem, R~Aqr and V1016 Cyg. For Mira, we repeated the observations of \cite{Sokoloski2010} but more accurately calibrated and at a much higher cadence, looking for the high frequency end of the power spectrum. Mira consists of a variable red giant Mira A, which is the prototype of Mira variable stars, and its companion Mira~B (aka VZ Ceti). Mira~A is a large amplitude variable with a period of about 331 days \citep{Mayer+2011}. The nature of Mira~B is still undetermined; with an effective temperature of about 10\,000K, it might be either an MS star or a WD. From the {\it Hubble Space Telescope} ({\it HST}) observations, \cite{Karovska+97} found that the separation between Mira~A and B is about 0.6$\arcsec$, which corresponds to a projected distance of $\sim$70 AU.\\
YGem is a semi-regular variable of spectral type M8 III which has a hot and strongly variable companion that has been detected by GALEX \citep{Sahai2011}. They found that the companion might have a blackbody temperature as high as 38,000K based on the near-to-far UV flux density ratio and atmosphere models by \cite{CK03}. \cite{Sahai2011} found that this UV emission is most likely to result from an accretion disk that surrounds an MS companion. This suggestion was based on the estimated emitting region's area that was found to be at least five times larger than the surface area of a WD. However, no classification has been confirmed for Y Gem's companion, yet. \\
R~Aqr is a well-studied and bright symbiotic system \citep{Whitlock+83} and it is well-known because of the two jets that extend about 1400AU from its centre. The system consists of a primary Mira (M7 III) with a pulsation period of about 387 days \citep{MatteiandAllen79} and a hot companion that is believed to be a magnetic WD with $M$ = 0.6 - 1M$_\odot$ \citep{Nichols+07}. The binary separation is about 0.55$\arcsec$ (200 AU; \citealt{Hollis+97}) and the distance to the system is about 363 pc (Hipparcos; \citealt{vanLeeuwen07}). \\
V1016 Cyg is considered to be a symbiotic nova, where the matter from the giant star accretes around the more compact companion. In 1964, V1016 Cyg underwent a slow nova eruption \citep{McCuskey65}, which confirms that the companion is a WD. UV observations show that the WD companion is very hot (150\,000K) and luminous (30\,000L$_\odot$) and its radiation ionises a fraction of the neutral wind from the giant Mira (e.g., \citealt{MN94}).\\
\indent The paper is divided into five sections. Section 2 describes the observations and the reduction. Results and discussion are given in sections 3 and 4, respectively. The last two sections summarises our comparisons and conclusions.

\begin{figure}
\includegraphics[width=1.05\linewidth]{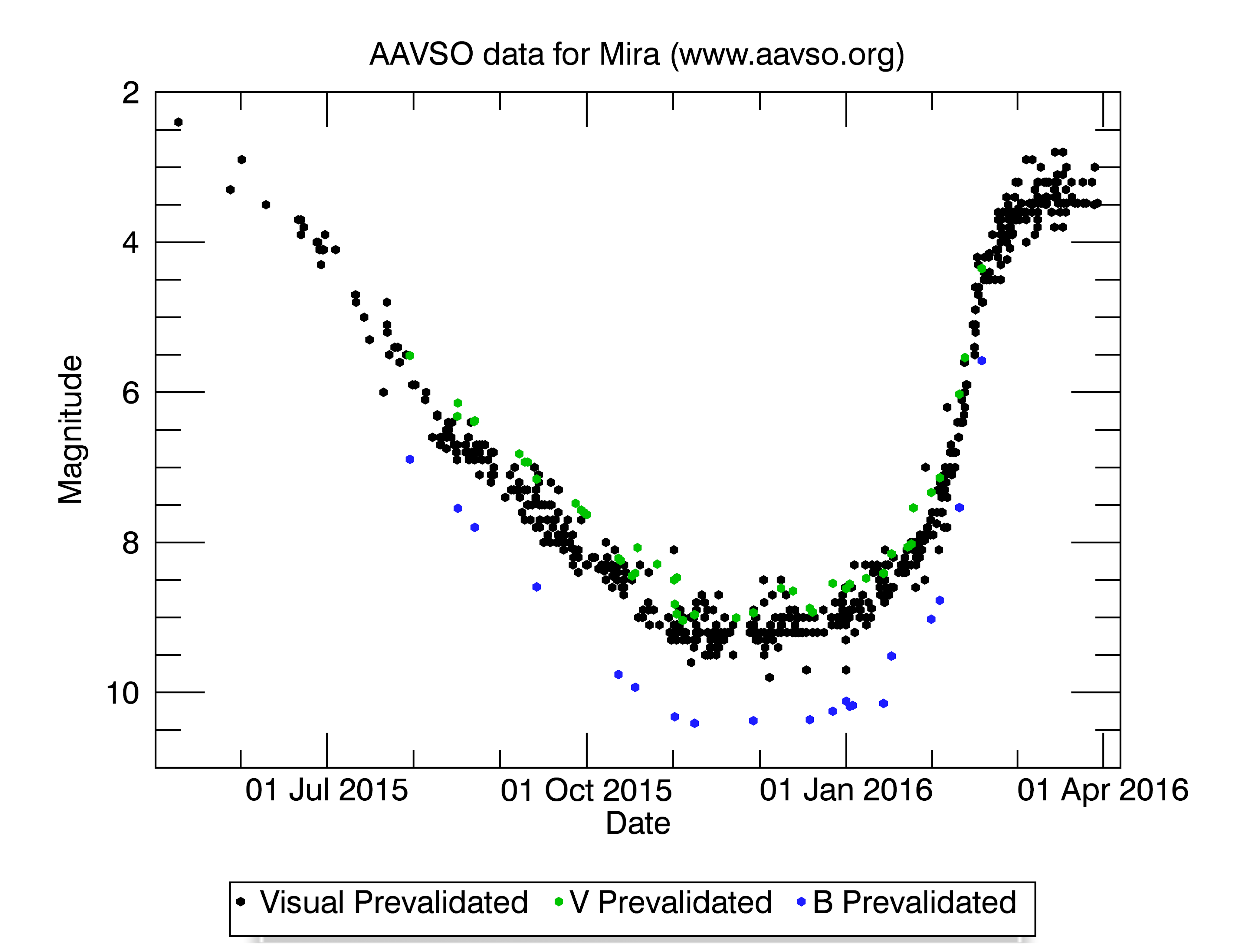}
\caption{The AAVSO light curve of Mira, in visual (black data points), V (green) and B (blue) filters. Our observations were taken at the end of September, when Mira was still about 1 magnitude brighter than at minimum. The flat part of the light curve during minimum is when Mira~B contributes most to the observed magnitude. }
\label{aavso}
\end{figure}

\begin{table*}
	\caption{Photometry and Comparison stars}
	\label{table:comp}
	\small
	\begin{minipage}[b]{1.0\hsize}\centering
		\begin{tabular}{lccclccccc}
			\hline
			\hline 
			\multirow{2}{*}{Target} & \multicolumn{3}{c}{Mag.} & \multirow{2}{*}{Comparison star} & RA & DEC & \multicolumn{3}{c}{Mag.} \\
			& $r'$ &  $g'$ & $u'$ & & (hh:mm:ss) & (dd:mm:ss) & $r'$ & $g'$ & $u'$  \\ \hline	
			V1016 Cyg   &  11.20 & 11.61 & 11.46 & TYC 3141-577-1         & 19:57:13.5  & +39:52:57.6   & 11.22\textsuperscript{1} & 12.22\textsuperscript{1} & 13.79\textsuperscript{2} \\
			& & & & USNO-A1.0 1275-13058156 &            19:57:04.9 & +39:49:14.4    &      &   &                \\
			R Aqr           & 7.57\textsuperscript{1} & 10.67\textsuperscript{1} &    &   &    & & & &            \\
			Mira             & 6.09\textsuperscript{1}   & 8.75 & 9.36   & HD 14411     &      02:19:28.5    & -02:57:57.5            &  8.82\textsuperscript{1} & 10.14\textsuperscript{1}&  12.57\textsuperscript{2}  \\
			Y Gem                &  & 9.99      &                & TYC 1369-542-1    &    07:40:58.2    & +20:24:07.6                     & 10.61\textsuperscript{1} &   11.13\textsuperscript{1}	& \\
			&   & 9.67 &  & TYC 1369-678-1  &  07:41:14.6 & +20:28:30.3  & 11.26\textsuperscript{1} & 12.65\textsuperscript{1}  &    \\   
			\hline
			\hline 
		\end{tabular}
	\end{minipage}%
	\begin{tablenotes}\footnotesize
		\item[*] (1) \cite{Henden+16}, (2) \cite{PD10}
		\item 
		\item
	\end{tablenotes}
\end{table*} 

\begin{figure*}
\includegraphics[trim=1cm 2cm 0cm 1.1cm, clip=true, width=0.39\textwidth, angle=-270]{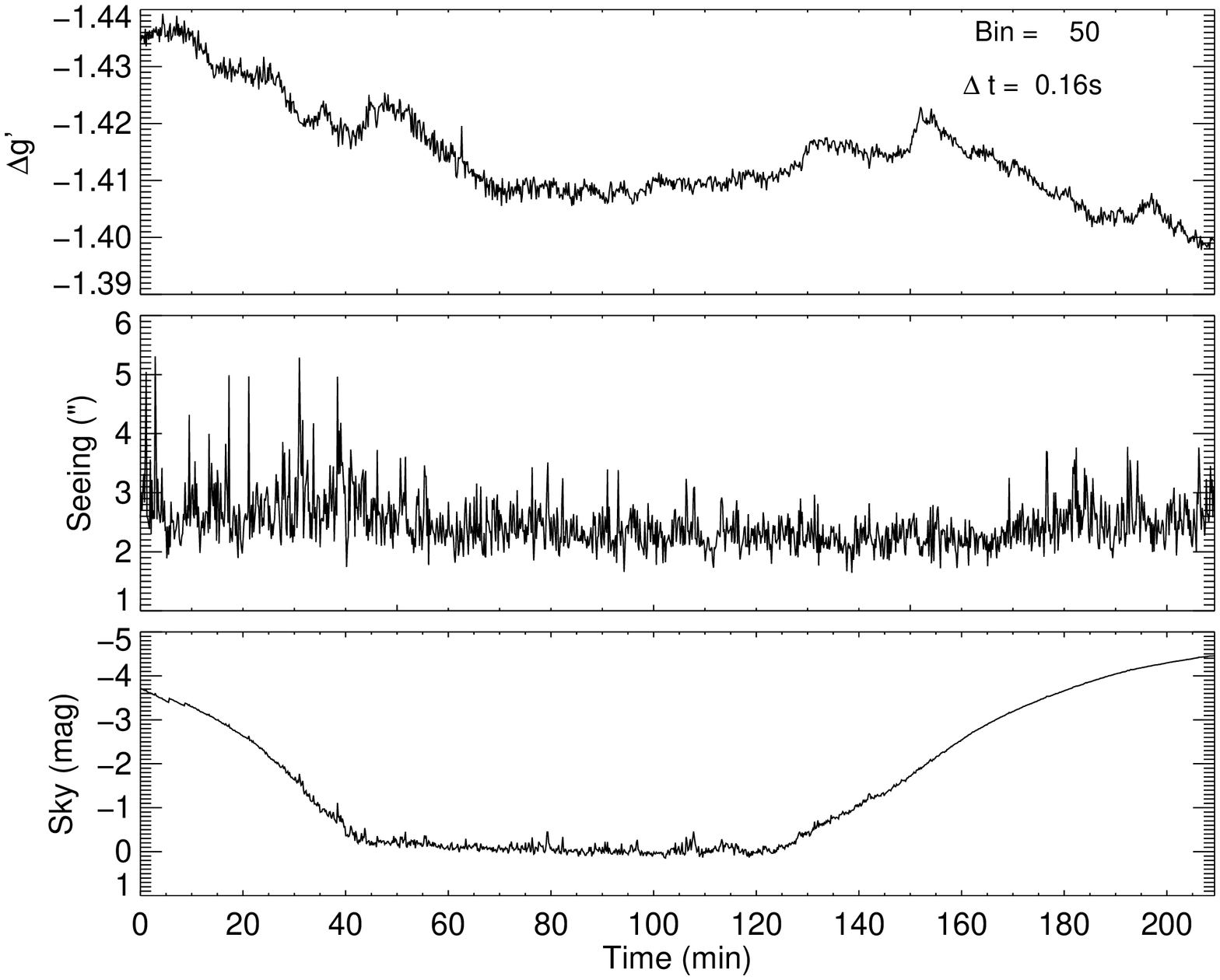}\quad\includegraphics[trim=1cm 2cm 0cm 1.1cm, clip=true, width=0.39\textwidth, angle=-270]{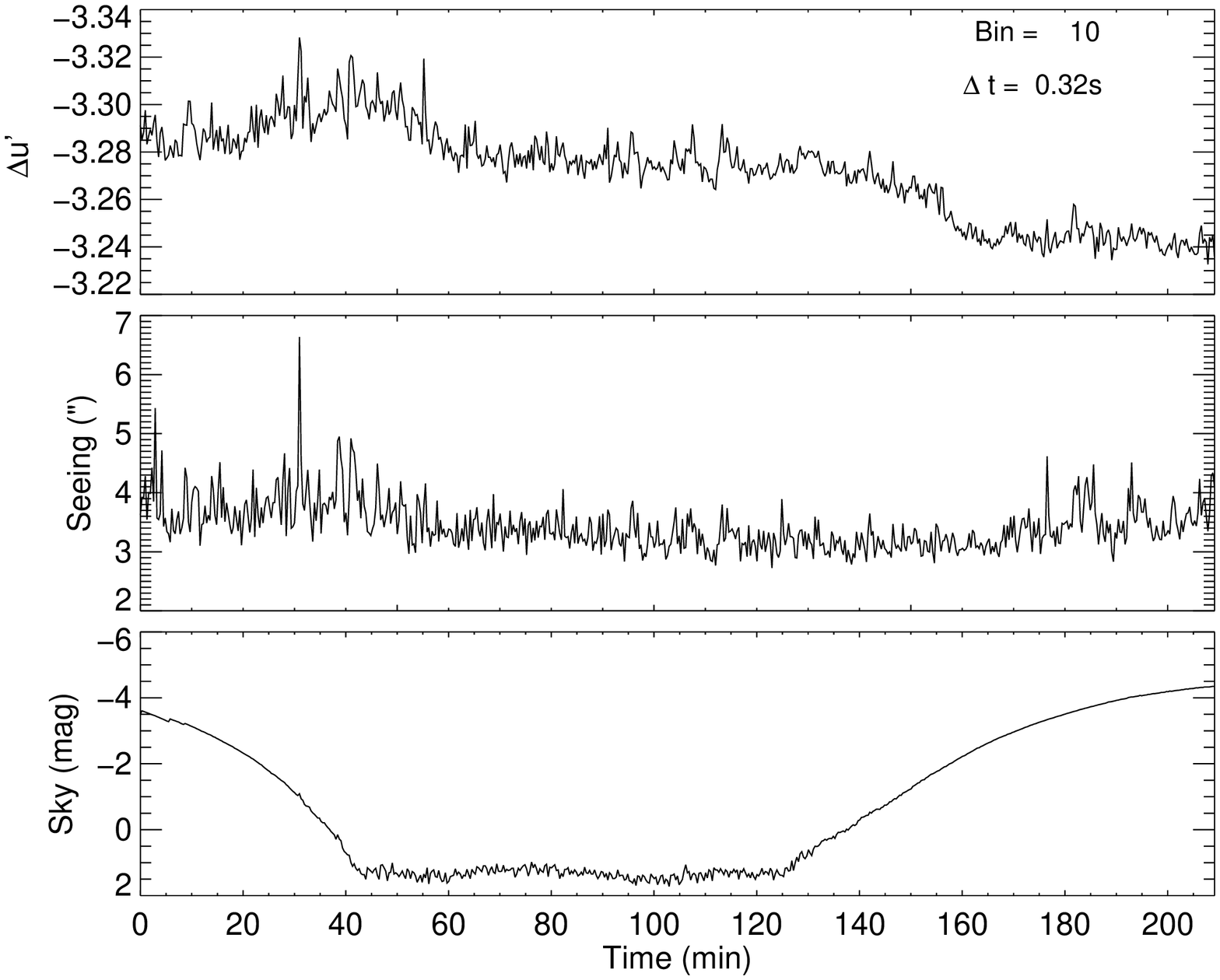}
\caption{(top) Differential light curves of Mira AB on the first night of observation in $g'$ (left) and $u'$ (right). The middle panel shows the image quality, set by the seeing and by the defocus of the telescope. The bottom panel shows the sky brightness within the aperture, in instrumental magnitude. The number of binned frames and the correspondent exposure time are shown on the top panel. The large variability in sky brightness is due to a lunar eclipse.}
\label{Mira1dm}
\end{figure*}

\begin{figure*}
\includegraphics[trim=1cm 2cm 0cm 1.1cm, clip=true, width=0.39\textwidth, angle=-270]{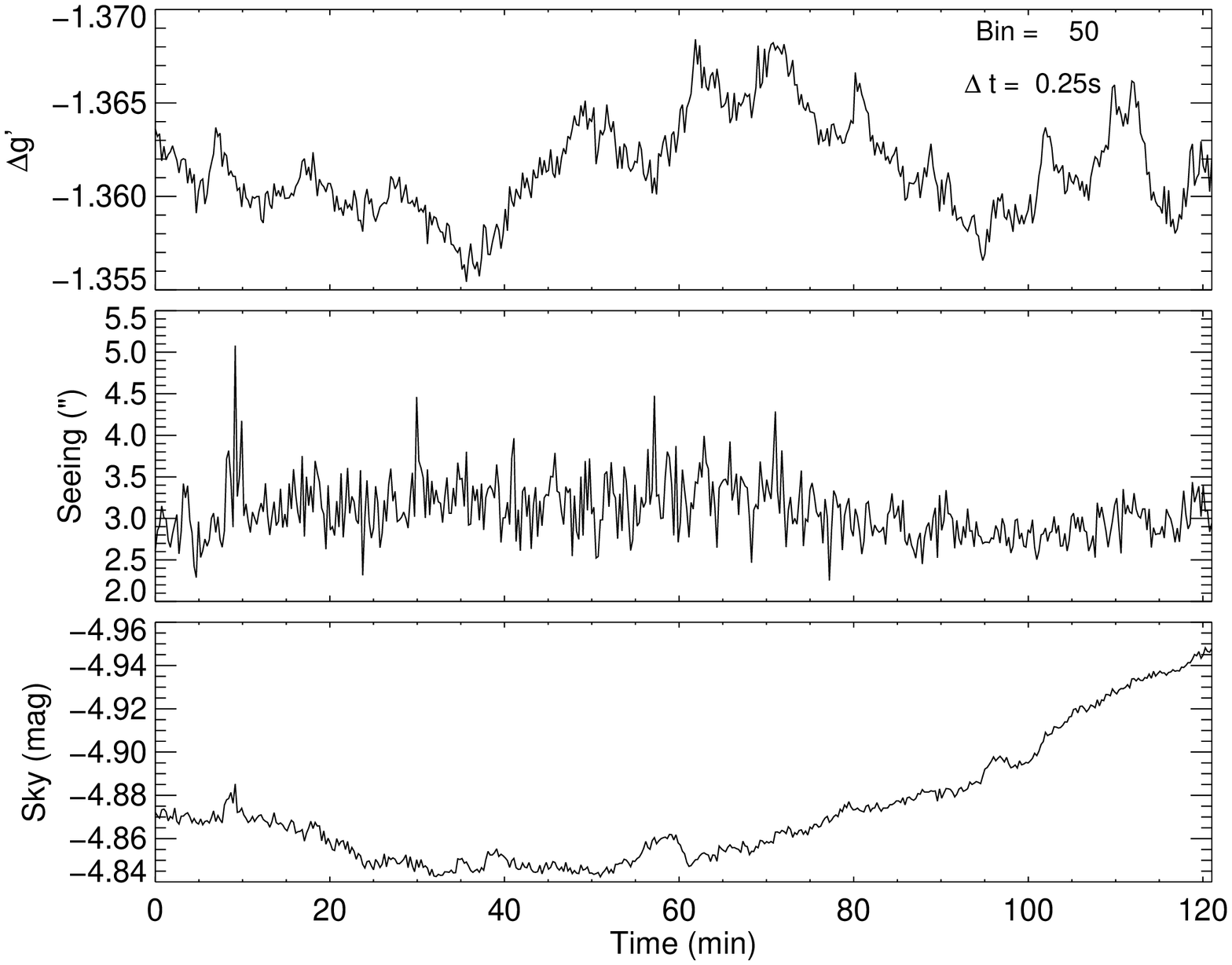}\quad\includegraphics[trim=1cm 2cm 0cm 1.1cm, clip=true, width=0.39\textwidth, angle=-270]{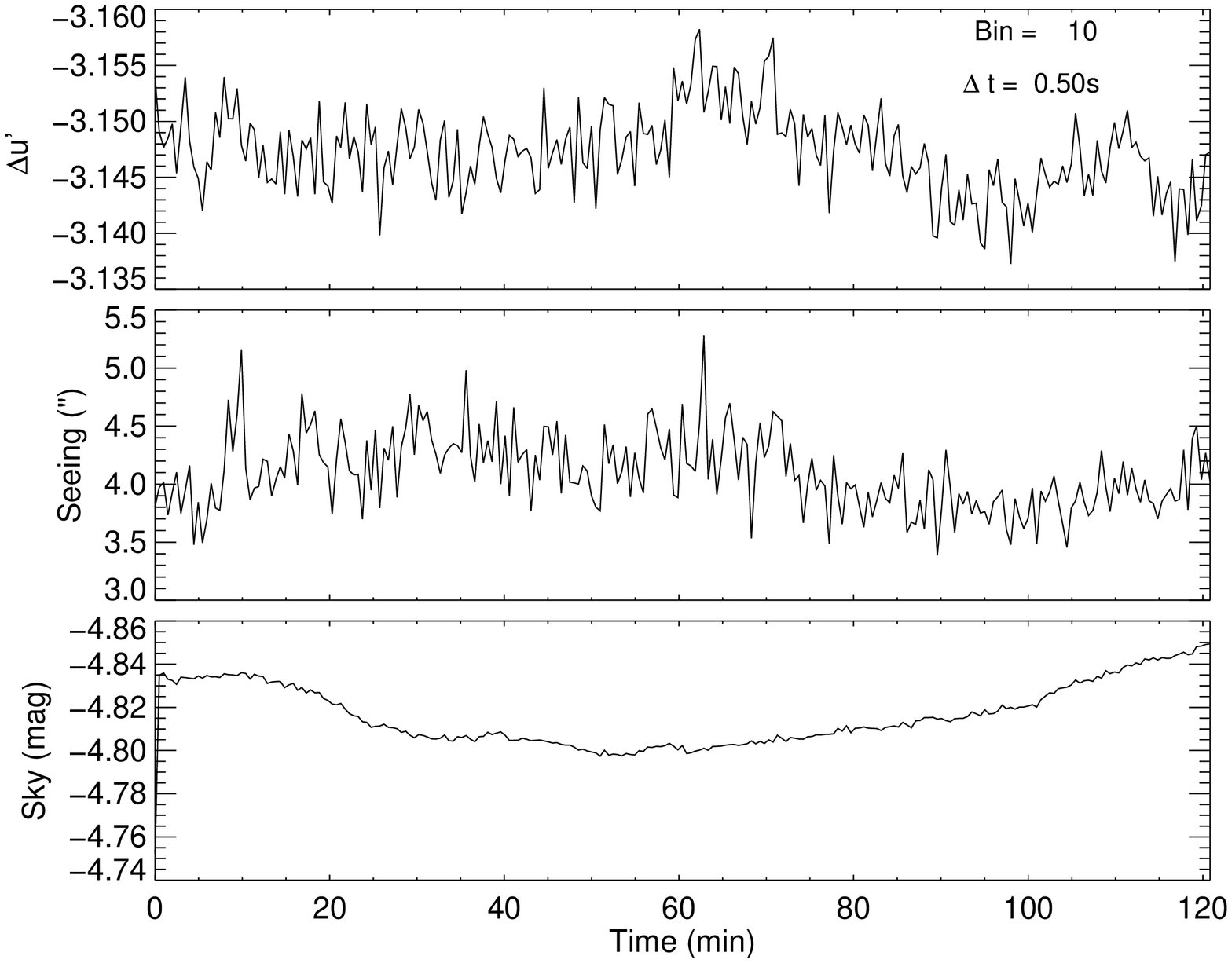}
\caption{Differential light curves from Mira AB for the second night of observations. Panels are as in Fig. \ref{Mira1dm}}
\label{Mira2dm}
\end{figure*}

\begin{figure}
\includegraphics[trim=0cm 2cm 0cm 1cm, clip=true, width=0.39\textwidth, angle=-270]{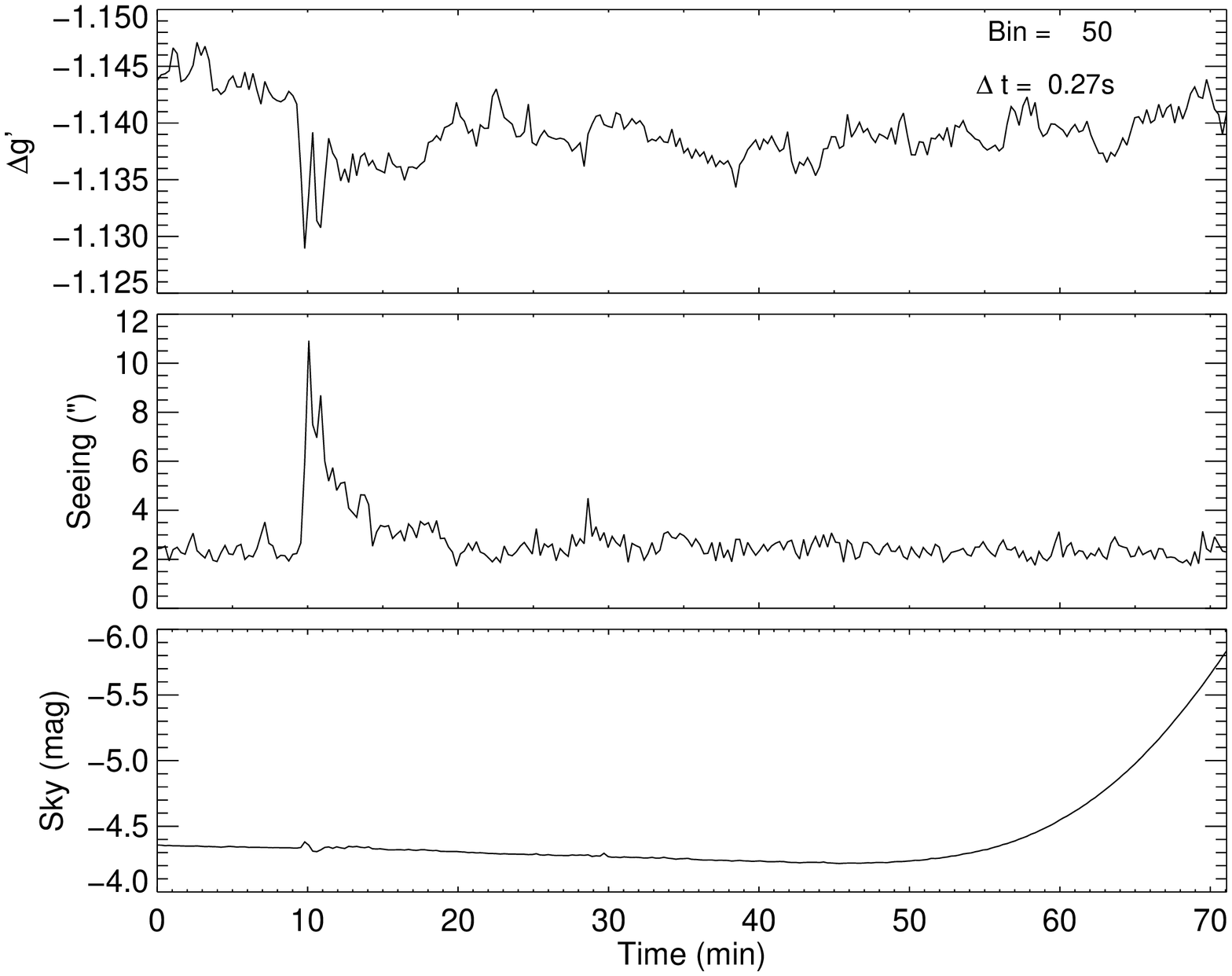}
\includegraphics[trim=0cm 2cm 0cm 1cm, clip=true, width=0.39\textwidth, angle=-270]{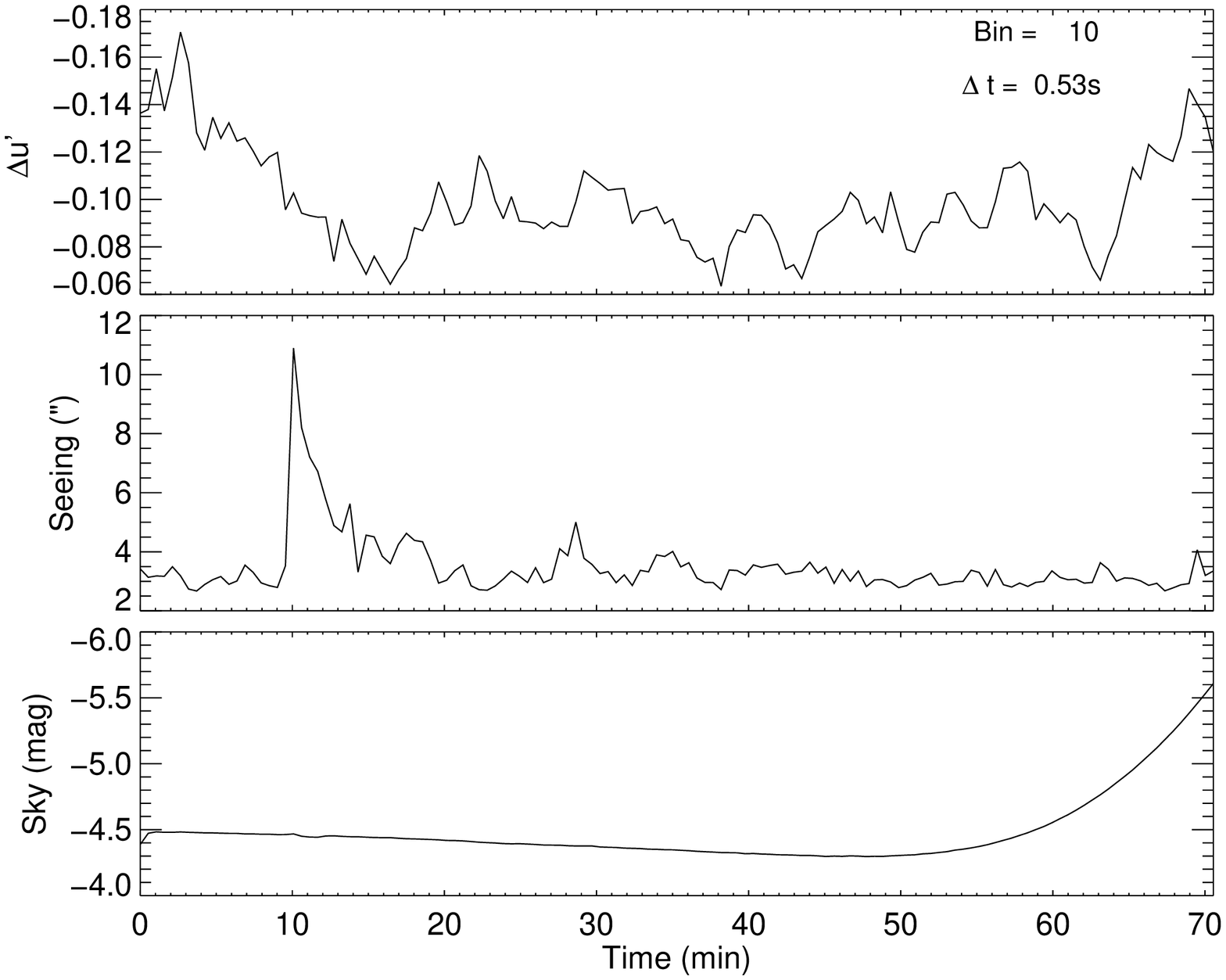}
\caption{Differential light curves for Y Gem and the comparison star TYC 1369-542-1 in $g'$ (top) and $u'$ (bottom) filters. Panels are as in Fig. \ref{Mira1dm}. }
\label{ygem-ref2}
\end{figure}	

\begin{figure}
\includegraphics[width=0.98\linewidth]{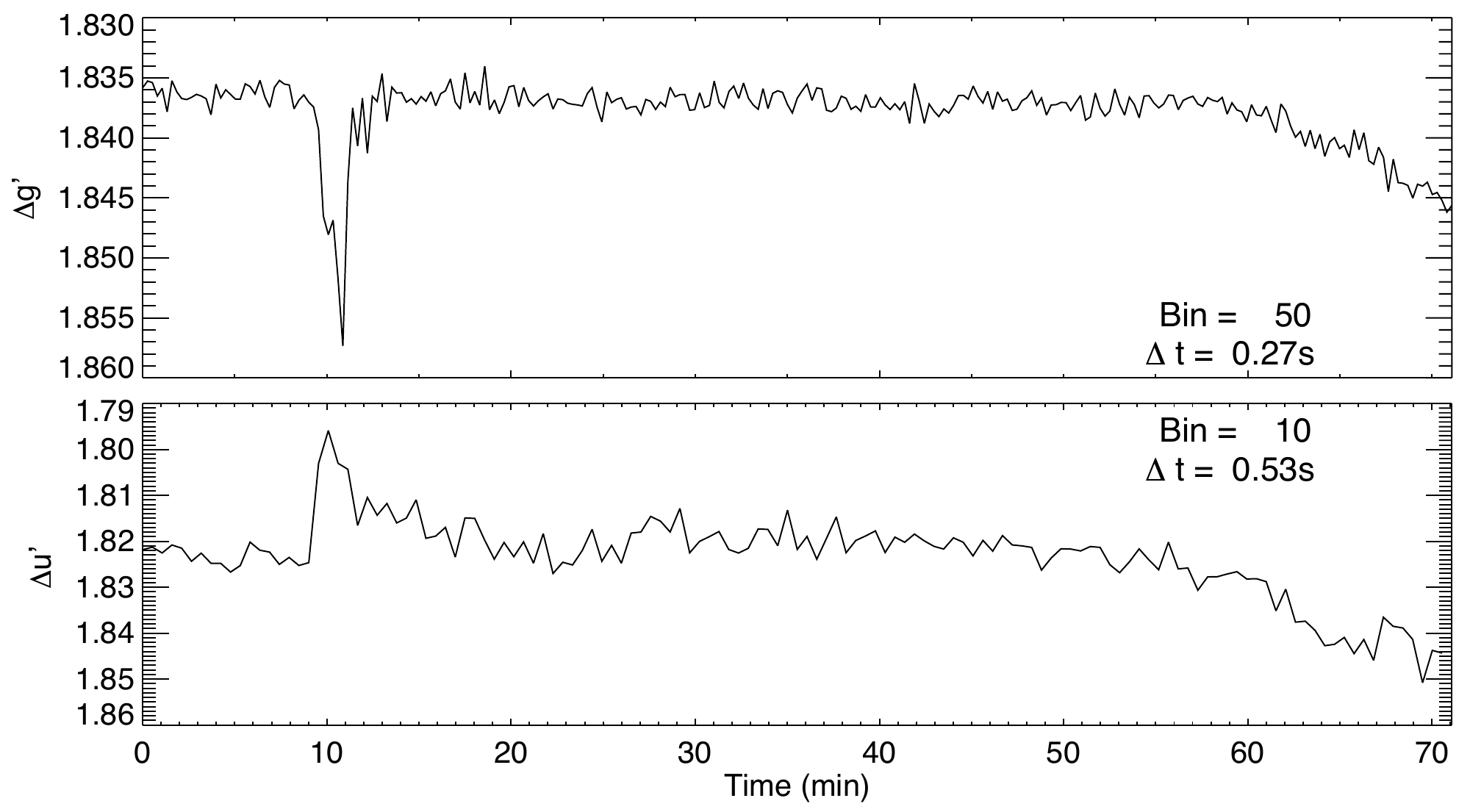}
\caption{Differential light curves between the two comparison stars of Y Gem. The first comparison star is TYC1369-542-1, the other comparison star is TYC 1369-678-1.}
\label{ygem-refs}
\end{figure}

\section{Observations and Data Reduction}
{All of the observations presented in this work were performed using ULTRACAM, a portable high-speed, triple-beam CCD camera mounted on the 4.2m William Herschel Telescope (\textit{WHT}) in La Palma, Spain. ULTRACAM (which stands for ULTRA-fast CAMera) uses the $ u' $$ g' $$ r' $$ i' $$ z' $ filter system defined by the Sloan Digital Sky Survey (SDSS, see e.g. \citealt{Smith2002}) and has the ability to observe in three colours simultaneously with a low readout noise ($\sim 3 e^{-}$) and short dead times (25 milliseconds) between exposures \citep{Dhillon2007}. It uses three 1024$\times$1024 CCD detectors with a platescale of 0.3\arcsec/pixel on the \textit{WHT} which gives a field of view of $\sim$300\arcsec. \\
Our objects were observed in $ u'$, $g'$ and $r'$ filters on two nights, starting from the evening of 27\textsuperscript{th} until the dawn of 29\textsuperscript{th} September 2015, during full Moon. A complete log of all observations is given in Table \ref{Observations}. During the first night there was a total lunar eclipse with a strong variation in the sky brightness.  Weather conditions were variable. During the first night there was occasional fog necessitating closure of the dome. Seeing was  between 1.5\arcsec - 2\arcsec, but it sometimes flared up to 7\arcsec. The second night showed much better conditions except for some large seeing spikes (up to $\sim$11\arcsec) which occured  about 10 minutes from the beginning of the observations of Y Gem.\\
The CCDs were windowed in order to reduce the exposure time and to avoid saturation of very bright targets. For Mira and Y Gem, the two brightest targets, it was also necessary to strongly defocus the telescope in order to avoid CCD saturation. However, we still found that the  $r'$ band was often saturated and consequently was not used. The two stars are considerably fainter in $u'$ and $g'$ and no saturation was seen in these filters. The sky was invariably much fainter than the stars. We used the fastest available frame rate, and binned the data to the required exposure time per point, typically 0.1 - 0.5 sec. \\
In each field, we selected two isolated comparison stars of similar brightness (within two magnitudes) to the target. In the case of Mira only one reference star was available within the field of view (see Table \ref{table:comp}). For V1016 Cyg, we found only one comparison star (TYC 3141-577-1) of a comparable  brightness to V1016 Cyg. The other comparison star is fainter. Unlike other targets, the field of V1016 Cyg is very crowded and, therefore, extra care was taken when choosing the sky annulus for the aperture photometry to avoid contamination from surrounding stars. \\
\indent Data reduction was performed using the ULTRACAM pipeline software\footnote[1]{http://deneb.astro.warwick.ac.uk/phsaap/software/ultracam/html/}. Bias frames were subtracted and then fields were divided by the flat field. The sky brightness was measured from the outer annulus, scaled to the number of pixels and subtracted from the measured flux.
Aperture photometry of all targets (except for R~Aqr) was performed using the variable extraction method where the radius of each aperture varies between frames as a multiple of the instantaneous seeing. The best aperture is normally around 1.5 times the full width at half maximum (FWHM) of the seeing profile \citep{Naylor98}. However, it tends to be larger for bright stars. Thus, we tried a range of scale factors (1.5, 2, and 3), then the scale factor that produces the best signal-to-noise ratio is chosen for the final reduction. The lowest factor was 1.5 and it was picked for the fainter target V1016 Cyg. The larger factors were used for the other brighter targets (2 for Mira on both nights, and 3 for Y Gem). Fixed aperture extraction was used for R~Aqr because one of the comparison stars was located close to the edge of the CCD window. The extreme brightness allowed for large apertures to be used.  For every detector integration time, photometry was obtained for the target and for the comparison stars, and the seeing was measured from the profiles, in each of the three filters. The data are presented as differential magnitudes between the target and each comparison star.

\section{Results}
\subsection{Mira}
For Mira, \cite{Sokoloski2010} reported rapid variability on a timescale of minutes to tens of minutes. As the variability is likely related to the companion, it is important to observe at a time when the brightness of the primary is near minimum. Our observations took place while Mira was approaching its minimum (Fig. \ref{aavso}). The AAVSO\footnote[2]{https://www.aavso.org} magnitude was $V\sim8.2$, which is one magnitude brighter than minimum, and $B-V\sim1$. The companion Mira~B has $V=9.5$. Mira~A and B are not separated in our images and therefore the photometry is the sum of their fluxes. As seen in our data, Mira faded by about 0.05 mag in $g'$ and by 0.1 mag in $u'$ between the two nights. This fading is consistent with the AAVSO light curve of Mira and represents its long-period variability. \\
Figures \ref{Mira1dm} and \ref{Mira2dm} show differential light curves of Mira and the comparison star HD 14411, on the two nights. HD 14411 is a K4-5III star, which formerly has been suspected to be variable. However, further observations have not confirmed its variable nature (e.g. \citealt{SS01}). In general, a K-type giant is expected to vary at a timescale much longer than our observations. The giant HD 14411 has a $g'$ magnitude of 10.14$\pm$0.18 \citep{Henden+16}, which gives us an estimate of Mira's magnitude that ranges from $g'$ = 8.78 - 8.72. about 0.5 magnitude fainter than the V magnitude in the AAVSO data. There is no $u'$ photometry published for HD 14411. \\
Compared to \cite{Sokoloski2010}, the current observations are at higher cadence giving better coverage of the high frequency end of the power spectrum, and benefit from simultaneous observations in multiple filters.\\
Mira was observed on two nights with different atmospheric conditions. During the first night, seeing (as set by the atmospheric seeing and telescope defocus) and sky brightness was variable, the latter because of the lunar eclipse. We plotted the differential magnitude against both parameters, but found no correlation for sky brightness. For the seeing, the calculated correlation coefficients show that a weak to moderate  correlations (ranging from 0.14 - 0.5) do exist except for V1016 Cyg that has a strong correlation coefficients in $u'$ ($\sim$0.6 - 0.7). However, this has no effect on our flickering measurements since it has been detected only in Mira and Y Gem and the correlation coefficient for these two targets are weak ($<$0.36 in all bands).
 The data (Fig. \ref{Mira1dm}) shows a slight fading over the full 3.5 hour in $g'$ by about 0.03 mag, with faster fluctuations. Variability is apparent at almost all timescales, up to an hour.  During the second night (Fig. \ref{Mira2dm}), when conditions were more stable, there was no secular fading but the faster fluctuations were again present, more prominent than on the first night. In addition, the $g'$ light curve of the second night is characterised by several microflares (a steep vertical rise in flux followed by a slow decay) of about 0.04 mag. The typical amplitude of the fast fluctuations is 0.005 mag on both nights. The rms noise in $g'$ on each data point is 0.01 mag on the first night and 0.003 mag on the second night (note that the integration times are slightly different). Figures \ref{Mira1dm} and \ref{Mira2dm}, corresponding to 0.004 and 0.002 mag per second. The rms noise was measured in a part of the light curve free from fast fluctuations. The peak-to-peak amplitude in $g'$ is 0.04 mag and 0.02 mag in the first and second night, respectively, but the value for the first night is dominated by the secular fading.  Without this, the peak-to-peak variability would be similar for the two nights.\\ 
One notable feature on the first night is a spike at 63 minutes after the beginning of observations, seen only in $g'$. It lasts for only one data point. We traced it in the data and the frame image does not show any artefact. We believe the spike results from cosmic rays that hit the $g'$ CCD. 
The $u'$ band shows a similar spike at approximately 55 minutes.} \\
The worst seeing ($\sim6.5\arcsec$) occurred about 30 minute after the start on the first night during a brief period. This corresponds to a peak in the $u'$ lasting for a few minutes. Otherwise there is no strong correlation between seeing and fluctuations in the differential magnitude.\\  
\indent The $u'$ curve is less complex and has less fluctuations compared to $g'$. The peak-to-peak amplitude, excluding the secular trend, is 0.04 mag and 0.02 mag during the first and second night. The rms noise is 0.01 mag per second and 0.003 mag per second respectively.
Comparing the $g'$ and $u'$ data, the secular trend on the first night is visible in both with a similar amplitude. But the fast fluctuations are much better seen in $g'$. Only the strongest fluctuations in $g'$ are mirrored in $u'$. The correlation is especially clear for the second half of the second night when the fluctuations were most pronounced. Narrow peaks in the fluctuations in $g'$ tend not to show a correspondence in $u'$. This may be due to the higher noise in the $u'$ data. \\

\subsection{Y Gem}
Y Gem is a semi-regular variable with a main period of 160 days. The variability is of low amplitude and it is not possible to define a well-determined minimum. The AAVSO visual magnitude varies between 9 and 10.5. Around the date of observation, $V=10$ was measured by AAVSO observers.
Only two reference stars of sufficient brightness in the ULTRACAM field were available for comparison. One is TYC1369-542-1 and the fainter one is TYC1369-678-1. Figure \ref{ygem-ref2} shows the differential light curve of Y Gem and the brightest comparison star TYC1369-542-1. The latter is a star that hasn't been classified nor studied yet, but some photometric measurements were available in publications (Table \ref{table:comp}).
Figure \ref{ygem-refs} shows the differential photometry between the two comparison stars. As neither appears rapidly variable, this gives a good indication of instrumental effects. The data show two significant deviations: a 0.02 mag excursion about 10 minutes into the observation, and a slow change by about 0.01 mag towards the end, starting first at $u'$. The latter coincides with the end of astronomical twilight and is due to an increasing contribution of the sky brightness especially for the fainter of the two comparisons. The first excursion was during a period of extreme seeing (11\arcsec\ was measured).  Apart from this, the differential magnitude between the two comparison stars is constant with an rms of 0.0010 mag/sec in $g'$ and 0.007 mag/sec in $u'$.\\
\indent The seeing and sky brightness are shown in Fig. \ref{ygem-ref2}. The seeing includes the effect of the defocussing of the telescope: the figure shows that the seeing degraded the image further only for the brief period around 10 minutes into the observation. In the analysis, data from this brief period was excluded. The $g'$ band shows the effect of the very poor seeing at 10 minutes. The slow change at the end seen in the second comparison star is not seen here as both stars are bright enough that the sky contribution remains negligible.  Excluding the period affected by seeing, the plot shows fluctuations with peak-to-peak amplitude of about 0.01 mag, or slightly larger if the first few minutes of data are included. This is three times larger than the  peak-to-peak variation seen between the two comparison stars. An apparent timescale for the fluctuations is of order 10 minutes. \\
\indent In the $u'$ band, the fluctuations are much larger, with peak-to-peak amplitude of 0.1 mag, or 0.06 mag excluding the beginning and end periods of the observation. This is far in excess of that shown between the two comparisons. 
Interestingly, the short-term variation is much larger in $u'$ than it is in $g'$. Assuming that the variability comes from the companion, this companion \citep{Sahai2008} must be bluer than Y Gem in ($u' - g'$). 
	
\subsection{V1016 Cyg and R Aqr}
\begin{figure}
\includegraphics[trim=0cm 2cm 0cm 1cm, clip=true, width=0.4\textwidth, angle=-270]{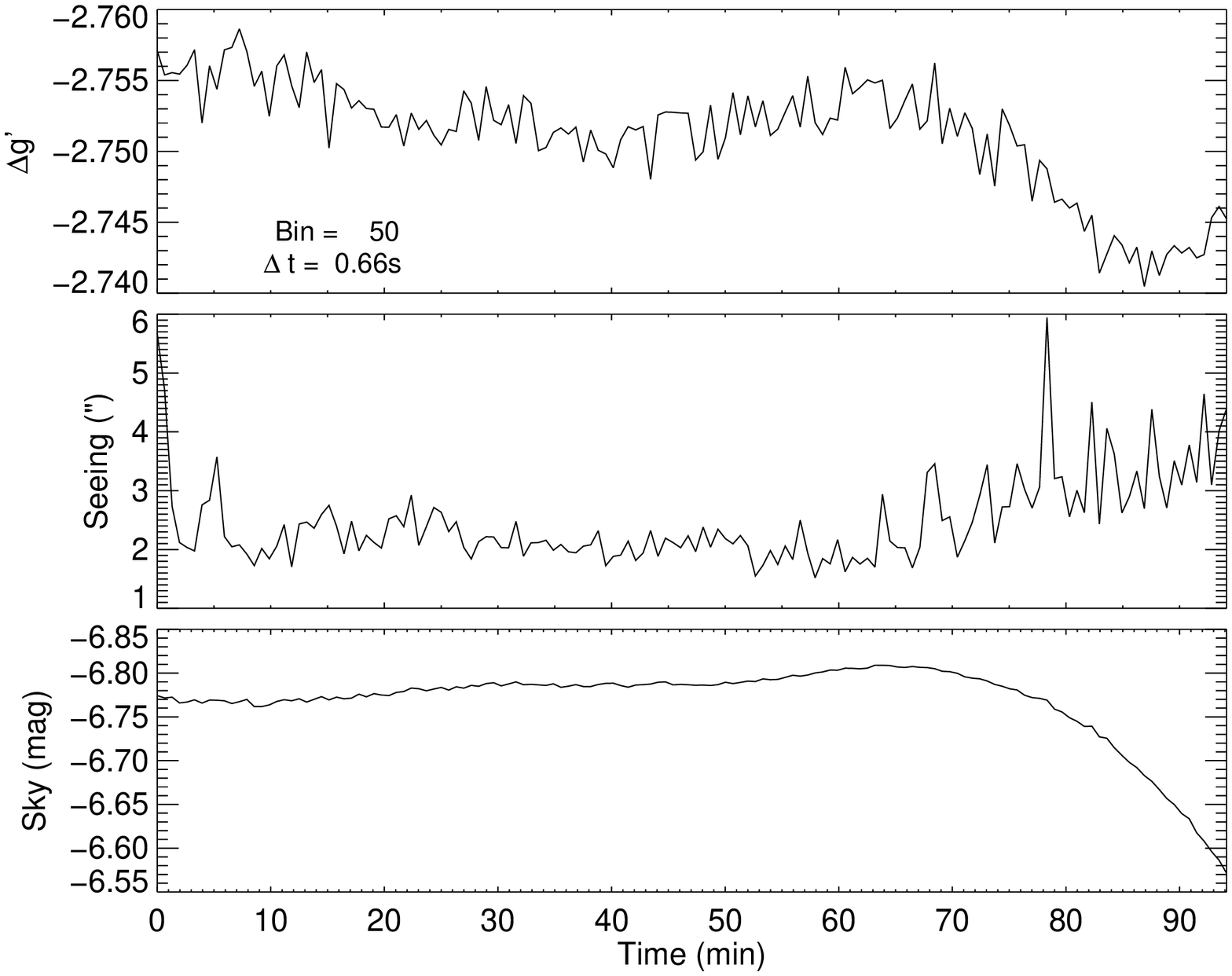} \includegraphics[trim=0cm 2cm 0cm 1cm, clip=true, width=0.4\textwidth, angle=-270]{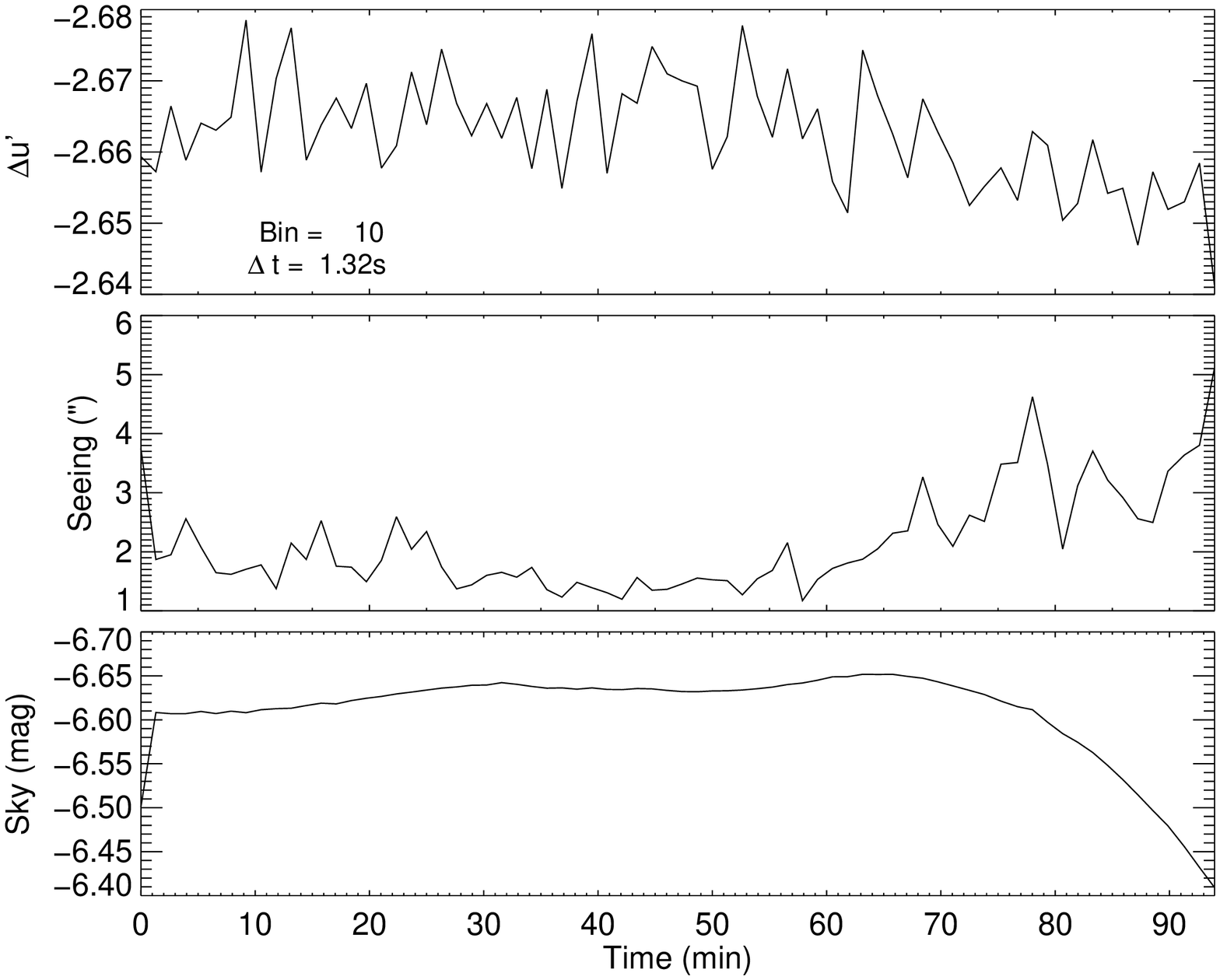}
\caption{Differential light curves for R~Aqr in $g'$ (top) and $u'$ (bottom) filters. Panels are as in Fig.\ref{Mira1dm}.} 
\label{raqr12dm}
\end{figure}

\begin{figure}
\includegraphics[trim=0cm 2cm 0cm 1cm, clip=true, width=0.4\textwidth, angle=-270]{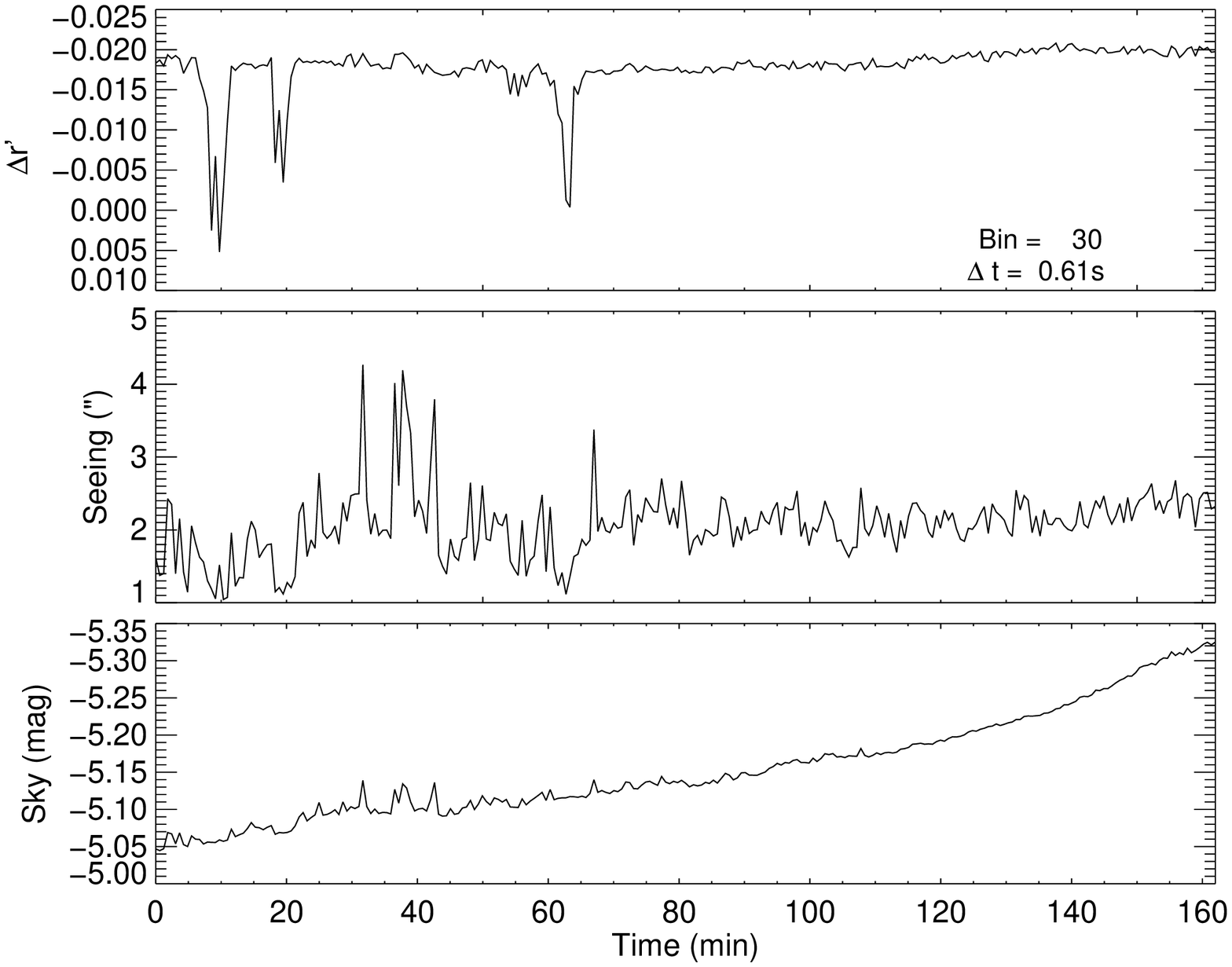}
\includegraphics[trim=0cm 2cm 0cm 1cm, clip=true, width=0.4\textwidth, angle=-270]{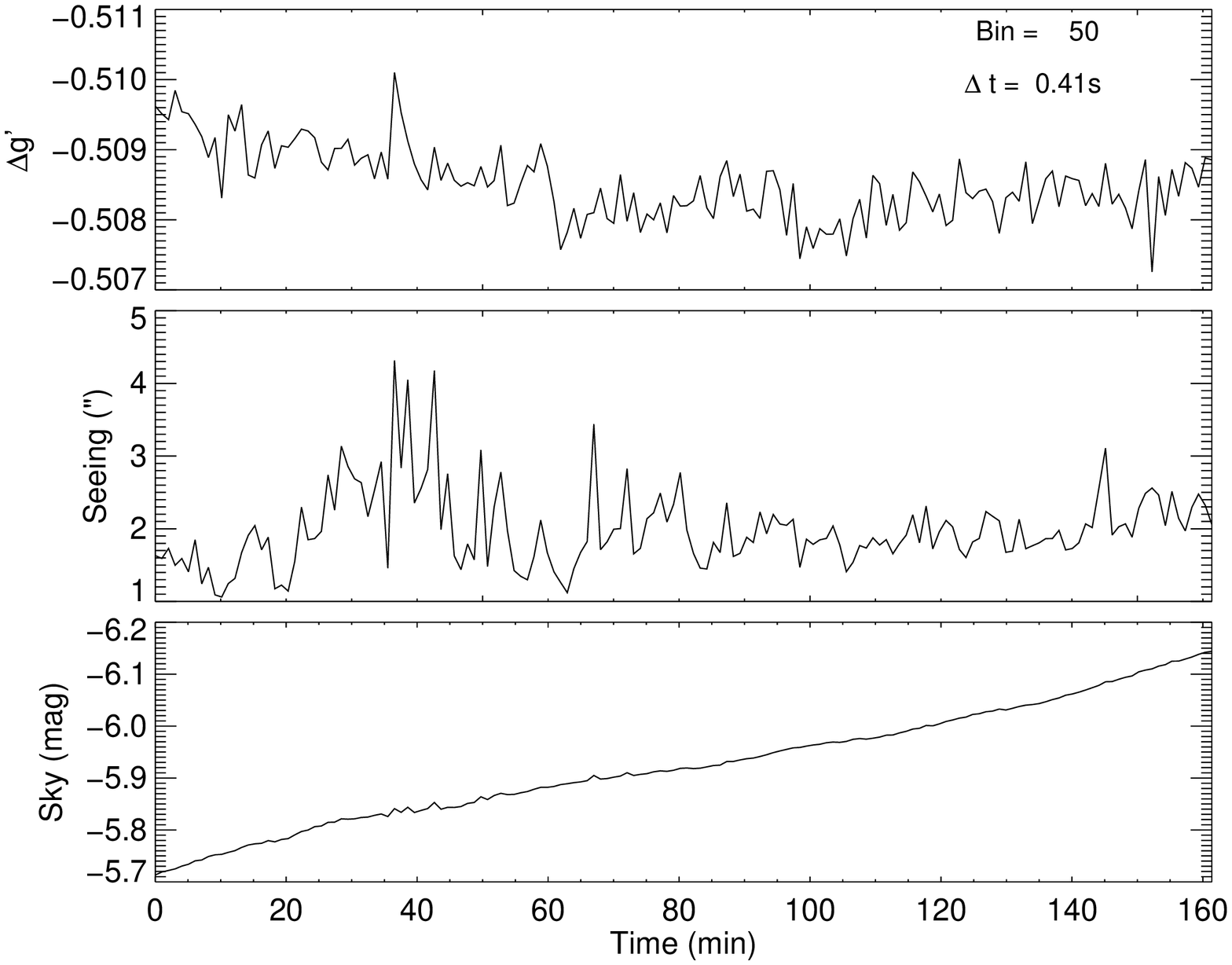}
\includegraphics[trim=0cm 2cm 0cm 1cm, clip=true, width=0.4\textwidth, angle=-270]{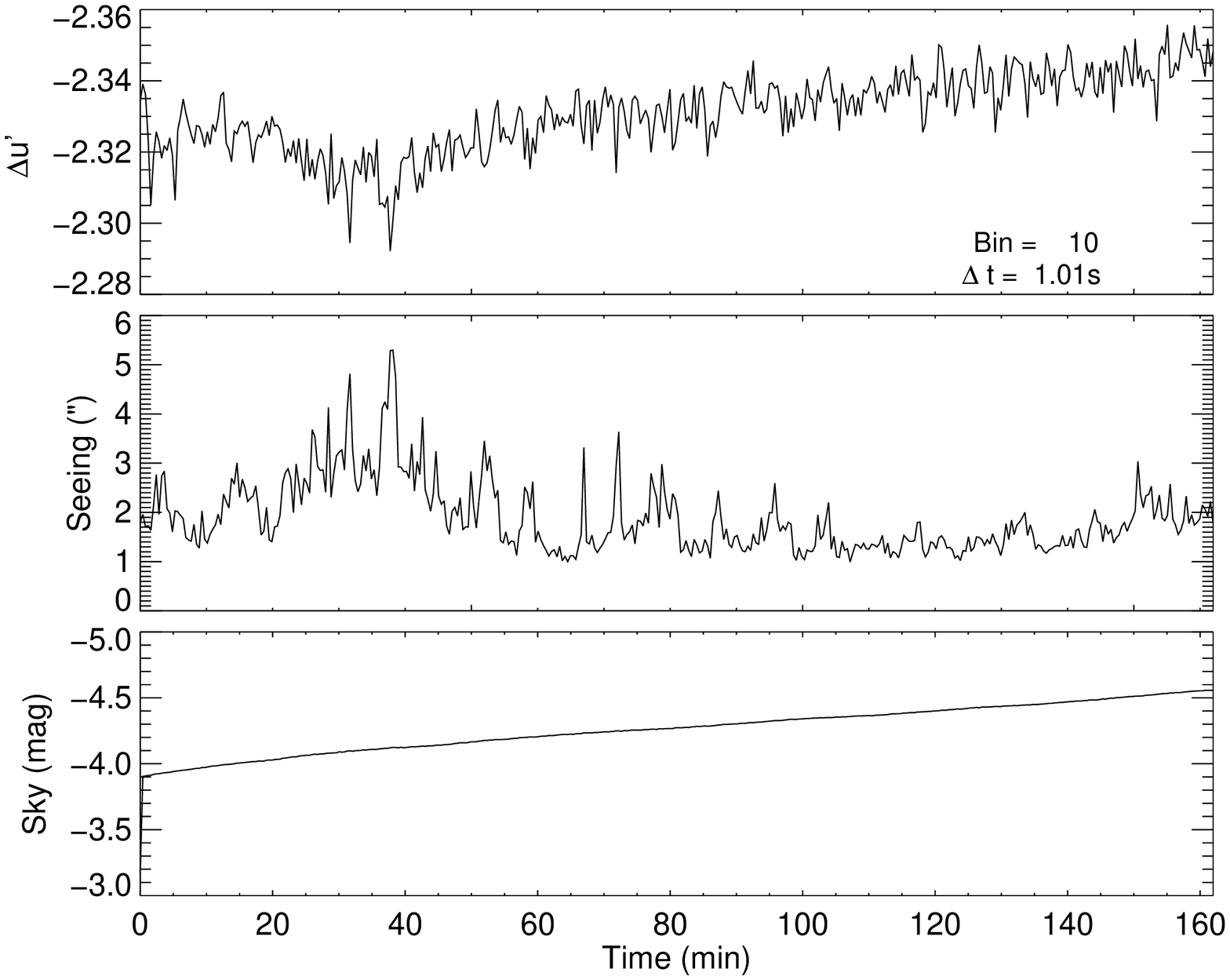}
\caption{Differential light curves for V1016 Cyg and the comparison star TYC 3141-577-1 in $r'$ (upper) $g'$ (middle) and $u'$ (bottom) filters. Panels are as in Fig. \ref{Mira1dm}.} 
\label{V1016Cygdm1}
\end{figure}

\begin{figure}
\includegraphics[trim=0cm 2cm 0cm 1cm, clip=true, width=0.4\textwidth, angle=-270]{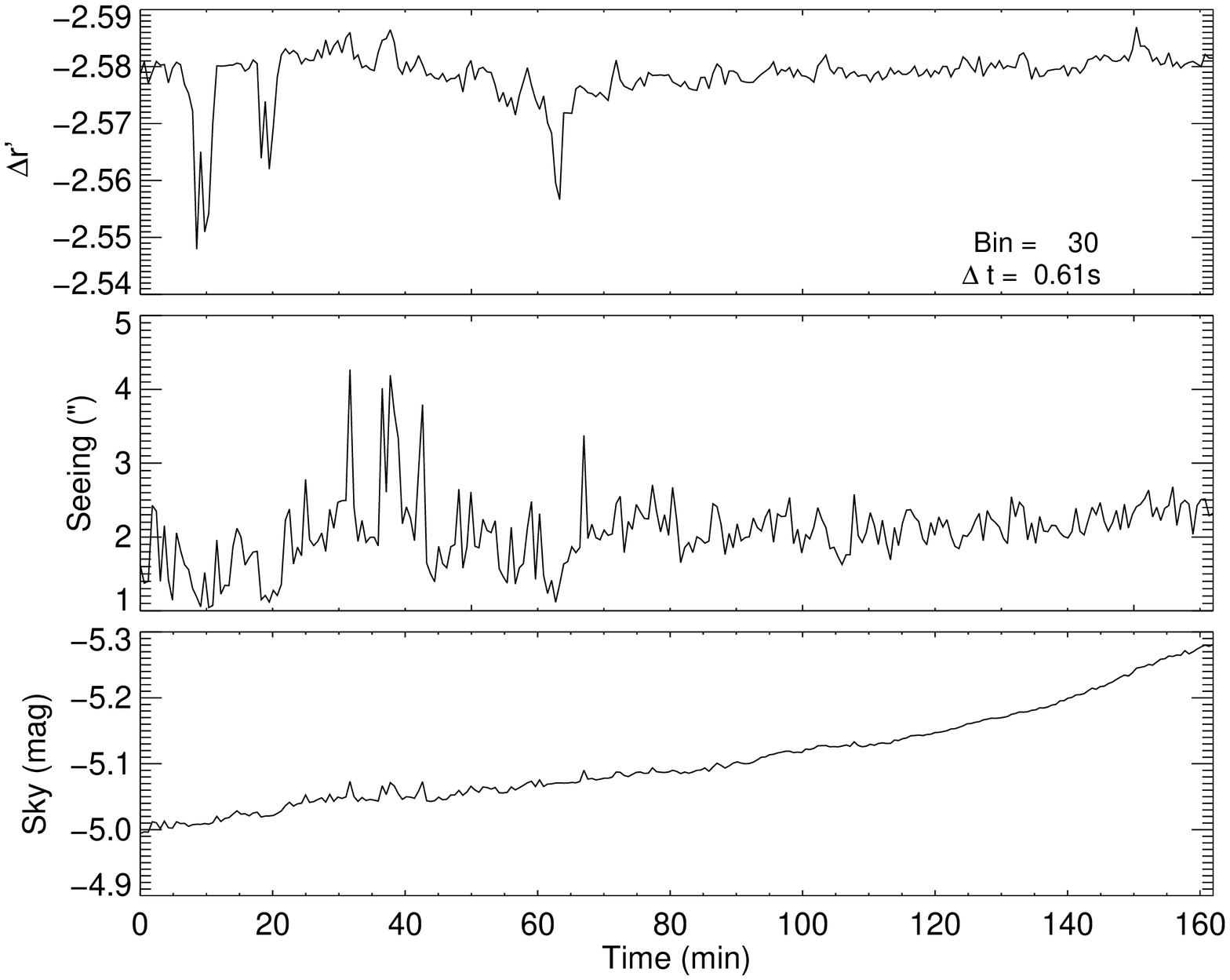}
\includegraphics[trim=0cm 2cm 0cm 1cm, clip=true, width=0.4\textwidth, angle=-270]{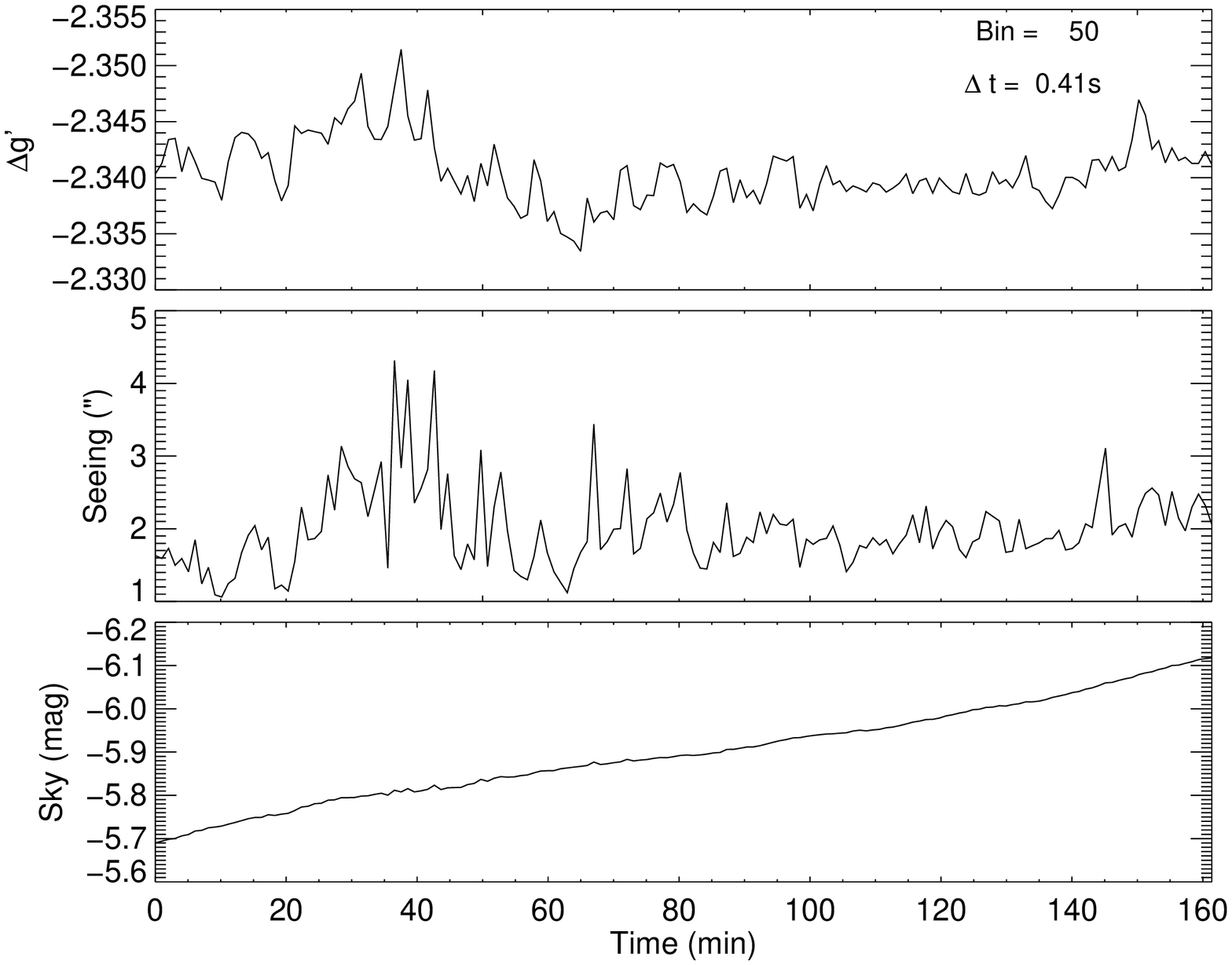}
\includegraphics[trim=0cm 2cm 0cm 1cm, clip=true, width=0.4\textwidth, angle=-270]{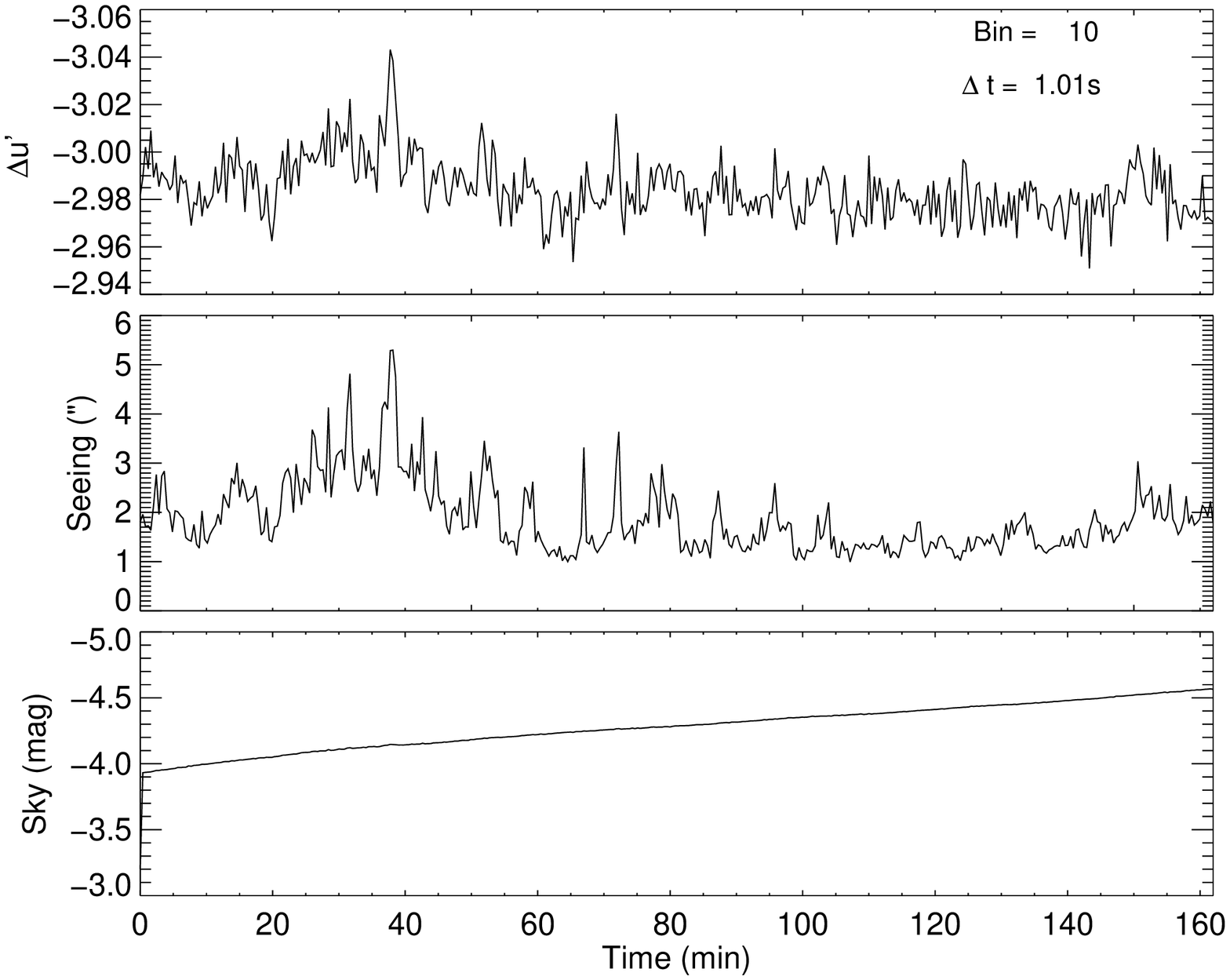}
\caption{Differential light curves for V1016 Cyg and the comparison star USNO A.10-1275-13058156 in $r'$ (upper) $g'$ (middle) and $u'$ (bottom) filters. Panels are as in Fig. \ref{Mira1dm}.} 
\label{V1016Cygdm2}
\end{figure}

\begin{figure}
\includegraphics[width=1.0\linewidth]{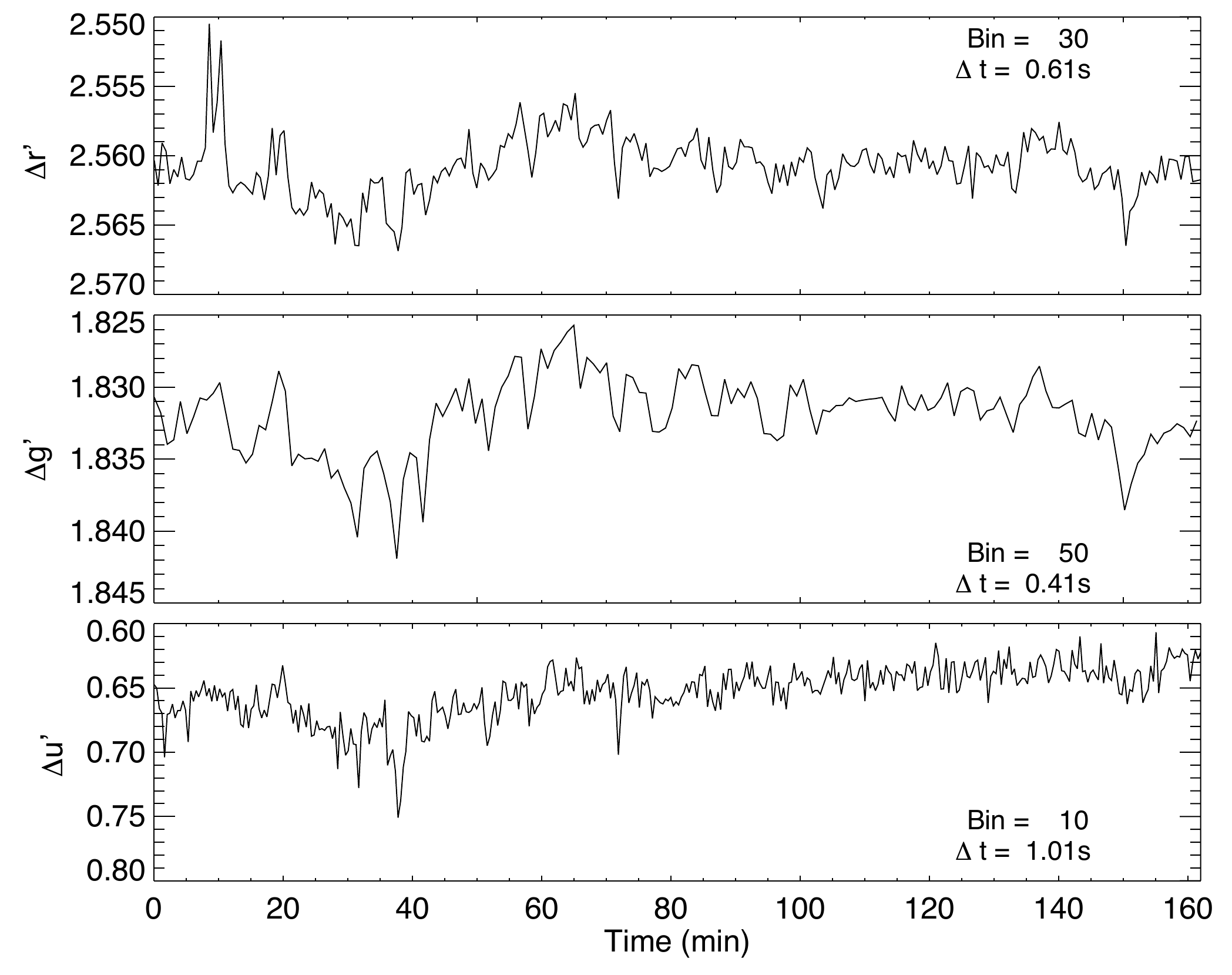}
\caption{Differential light curves in $r'$, $g'$ and $u'$ for the two comparison stars of V1016 Cyg: USNO-A1.0-1275-13058156 and TYC3141-577-1} 			
\label{V1016Cyg-refs}
\end{figure}

Both stars were observed on 27\textsuperscript{th}\ September 2015. 
For R Aqr, the two comparison stars could not be identified in photometric catalogues. After about 70 minutes of observations, the sky brightness had dropped because of the beginning of the lunar eclipse (by $\sim$ 0.3mag.). Seeing increased during the last 20 minutes of observations from 2$\arcsec$ to 5$\arcsec$. Good observations lasted only for 30 minutes as seen in Fig. \ref{raqr12dm}. \\
\indent V1016 Cyg is the only star that was not saturated in $r'$ CCD (Figs. \ref{V1016Cygdm1} and \ref{V1016Cygdm2}) Two field stars were chosen for comparison (Table \ref{table:comp}). TYC 3141-577-1 has a comparable brightness to V1016 Cyg, while the other comparison star (USNO-A1.0 1275-13058156) is much fainter. The light curve in $r'$ is characterised by three drops. We believe the first two result from systematic issues and not related to V1016 Cyg as they are present in the comparison stars' light curve too (Fig. \ref{V1016Cyg-refs}). The third drop does not appear in Fig. \ref{V1016Cyg-refs}. We traced it through corresponding frames and no obvious cause could be identified. Seeing started poor ($\sim$4\arcsec) and became better after that. The reflection of bad seeing could be seen in all the three light curves.\\
\indent For both V1016 Cyg and R~Aqr, no fluctuations could be seen in their light curves. R~Aqr was observed very close to the Moon and the sky brightness affected the photometry, as can be seen from the change near the end during the onset of the lunar eclipse (Fig. \ref{raqr12dm}). The rms noise was 0.01 mag/sec in $u'$  and 0.001 mag/sec in $g'$  The second reference star was fainter (and barely visible in $u'$) and gave higher noise so it wasn't used. For V1016 Cyg, some variation arose from seeing variation and sky brightness, and this was also seen in the differential magnitude between the two reference stars (Fig. \ref{V1016Cyg-refs}). The rms noise was 0.01 mag/sec in $u'$ and 0.001 mag/sec in $g'$. 

\begin{figure}
\includegraphics[width=1.0\linewidth]{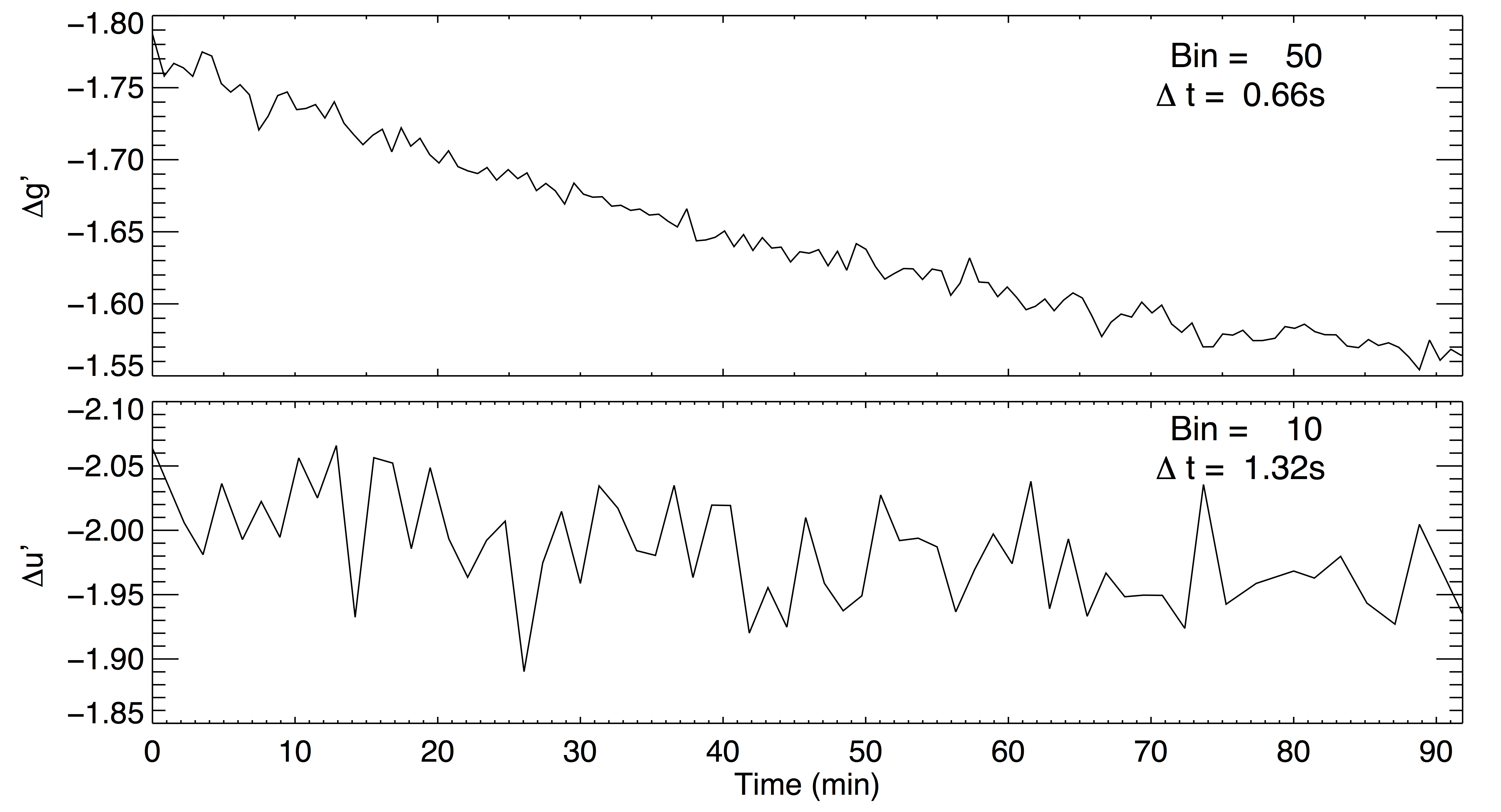}
\caption{Differential light curves between the two comparison stars of R~Aqr. The two comparison stars could not be identified.}
\label{raqr-refs}
\end{figure}

\section{Discussion}
\subsection{Power spectra}\label{section:power.spectra}
In all observations, our data points are not uniformly spaced, therefore Lomb-Scargle periodogram (\citealt{Lomb76}; \citealt{Scargle82}) was used in order to generate the power density spectrum (PDS). The resulted PDS was normalised using the method of \cite{Miyamoto+91}. In this normalisation, according to Parseval's theorem, the square root of the integrated periodogram over a range of frequencies is equal to the fractional rms variation over the same range. For sake of appearance, especially at higher frequencies, spectra were smoothed slightly (by a factor of 10), to a time resolution of approximately 1 sec.\\
\indent The power spectra for Mira and Y Gem are shown in Fig. \ref{power_mira2} - \ref{power_ygem-TYC6}, while the power spectra for R~Aqr and V1016 Cyg are given in the appendices (\ref{power_raqr_left}, \ref{Power_V1016Cyg1} and \ref{power_V1016Cygdm2}, respectively). For Mira, we had only one comparison star but two nights of observations. Power spectra for the two nights are shown separately. For Y Gem, there were two comparison stars: we show the power spectra using the brightest one, and for comparison the power spectrum of one comparison star against the other. For R~Aqr, there was only one comparison star in $u'$. For V1016 Cyg, we show the power spectra for the two comparison stars and for the comparison stars against each other.\\
\indent The power spectra show the frequency range from $2\times 10^{-5}$ Hz to $5.17\times 10^{-2}$ Hz. However, our data sensitivity covers only half of the total length of the observations, which means that the actual lowest frequency ranges from $1.6\times10^{-4}$ Hz (Mira, first night), to $5\times 10^{-4}$ Hz (Y Gem). The turn-overs seen at lower frequencies are not real. At the highest frequencies, our time resolution allows for variations to be detected up to 1 Hz but in practice we did not look for variations faster than 0.1Hz.\\
\indent From the power spectra we can extract the rms variation and the slope. The rms is obtained as the square root of the integrated power spectrum between $10^{-3}$ to $10^{-2}$ Hz. We picked this frequency range for two reasons: 1) it is the area where we have the best coverage; 2) it is the only area where we can compare our findings to those of \cite{BarrosPhD}, who estimated rms variations in this frequency range for cataclysmic variable (CVs). For Mira, we also compare rms variability in this region to rms percentage calculated by \cite{Sokoloski2010}. The slope is measured from a power law fit over the same frequency range. 
We could not find any evidence for periodicity in all our observations. Table \ref{power spectra} summarises the properties of power spectra.\\
We inspect the effect of seeing and sky brightness on power spectra. We found that seeing is not a source of fluctuations since its power spectra are almost flat. A similar result was found earlier by decorrelating seeing and light curves. On the other hand, sky brightness shows noticeable slope in its power spectra. However, this has no effect on our targets because they are very bright objects, except for V1016 Cyg, which is the fainter target.\\
  
\subsection{Mira}
Over five nights of observations during Mira A's minimum light, \cite{Sokoloski2010} found evidence of flickering with amplitudes of $\sim$ 0.2 mag on timescales of 1-10 minutes, which is significantly greater than that expected from an accreting MS star, but consistent with that of accreting WDs in CVs. Thus, they argue for the WD nature of Mira~B. In the UV, \cite{Warner72} found variation of order 0.008 mag, which is comparable with our findings. \\ 
From our observations of Mira, we can make an estimate for the fraction of the flux that can be attributed to the companion. We observed Mira when it was about 1 mag above its minimum (a small change in magnitude between the two nights can be attributed to its long period variation). Our measurements, which were taken in the Vega-based system, are $u^\prime = 9.36$ and $g^\prime = 8.75$ mag for the combined light of Mira~A and B.\\
Optical and UV photometry for Mira B were obtained from the ({\it HST}) study by \cite{Karovska+97} in the  AB-based system. For Mira~B, they measured a value of 11.32 at wavelengths corresponding to $B$, $g'$ and $V$ together , and 10.23 at wavelengths corresponding to $U$. 
We converted their values from AB- to Vega-based magnitude system using relations in \cite{FG94}, and we find that Mira B has $B = 11.48$, $V = 11.36$.\\
There is no tabulated conversion for $U$ from AB- to Vega-based system in \cite{FG94}. The standard Johnson $U$ band has a zero point of 1884 Jy, and the AB-system is defined as a zero point of 3631 Jy. Hence, we have 
$U_{Johnson} = U_{AB} - 0.71$. \\
We find that for an idealised Vega-like star (with $[\rm{Fe/H}]=~0, \, [\alpha/Fe]=~0, \, \rm log({\it g})\sim4.5, T=9600 K$) there will be approximately 22\% less flux in $u'$ than $U$. Therefore, Mira~B has
$u'_{Vega}$ = 9.8. We calculated that, during our observations, Mira~A has $g'$ = 9.77 and $u'$=10.55. \\
\indent The rough estimate of contributions from Mira~B is 8\%\ of the flux in $g'$ and 70\%\ in $u'$.
The fractional variabilities from Mira PDSs are given in Table \ref{power spectra}. Correcting for the flux coming from Mira A, and assuming all flickering power is due to the companion, the rms on Mira~B becomes 2.4\% in $g'$ and 0.25\% in $u'$ observations of the first night, and 8.8\% and 0.7\% in $g'$ and $u'$ observations of the second night. \\
\cite{Sokoloski2010} found an average value for rms variation in the $B$-band from Mira~B between $10^{-3}$ to $10^{-2}$ Hz of about 3\%, which is in agreement with \cite{BarrosPhD}, who found fractional rms variability for CVs of 2.4\% in $g'$ band and 4.5\% in the $u'$ band. 

\subsection{Y Gem, R Aqr and V1016 Cyg}
High UV emission from Y Gem had been detected by \cite{Sahai2011}, who argued that this emission is most likely resulting from an accretion disk around an MS companion. \\
We measure the Y Gem system to be $g^\prime$ = 9.99 or 9.67 mag, depending on whether TYC 1369-542-1 or TYC 1369-678-1 is used as the comparison star. No optical observations are available for the companion of Y Gem, but scaling from the UV flux densities from \cite{Sahai2011}, the companion wouldn't be brighter than $g'$=13.2, which would mean that the companion only contributes 4.4\%\ of the total flux in $g'$. However, this should be viewed with caution as the companion varies in UV brightness and its SED is not well known.\\
The light curve of Y Gem shows a clear signal of short-term fluctuations, as discussed before. The typical peak-to-peak fluctuations over short timescales of 0.005 mag in $g'$ and 0.05 mag in $u'$. That corresponds to a variability in the companion of order 20\%\ in $g'$, assuming $g'$ of the companion is 13.2.\\
\indent No significant flickering was seen for R~Aqr. Typical variations are around 0.01 mag in $u'$ and 0.002 mag in $g'$; the lack of trend suggests this is the noise level in the data. The power spectrum shows flat (white) noise at frequencies larger than 1 mHz (Fig. \ref{power_raqr_left}). R~Aqr has a companion, believed to be a WD, and an orbital distance of 200 AU. This is closer than in the Mira system, and the companion is known to be accreting because of the jets emanating from it, but this does not cause observable flickering.\\
\indent V1016 Cyg  also has a featureless light curve without clear indications for flickering. During the best conditions, the fluctuations on the differential magnitudes are of order 0.002 mag in $g'$ and $r'$, and a little  noisier in $u'$ (Figures \ref{Power_V1016Cyg1} and \ref{power_V1016Cygdm2}). The power spectrum shows white noise over the relevant frequency range. Although a known binary with a history of eruptions, there is no notable flickering at the time of observation.

\subsection{Classification of Mira~B}
\begin{figure*}
	\includegraphics[trim=0cm 2cm 0cm 1cm, clip=true, width=0.3\textwidth, angle=-270]{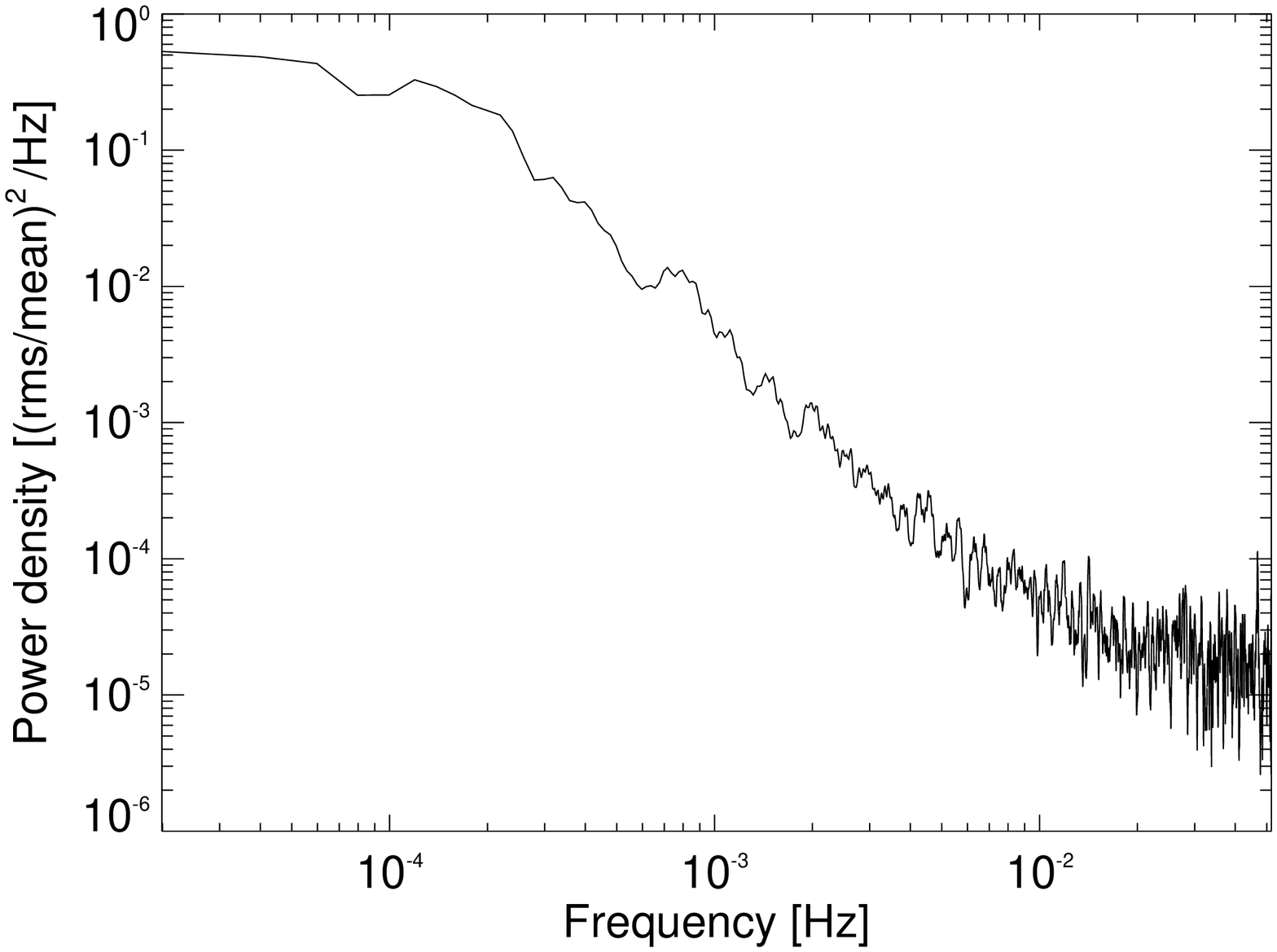}\qquad\includegraphics[trim=0cm 2cm 0cm 1cm, clip=true, width=0.3\textwidth, angle=-270]{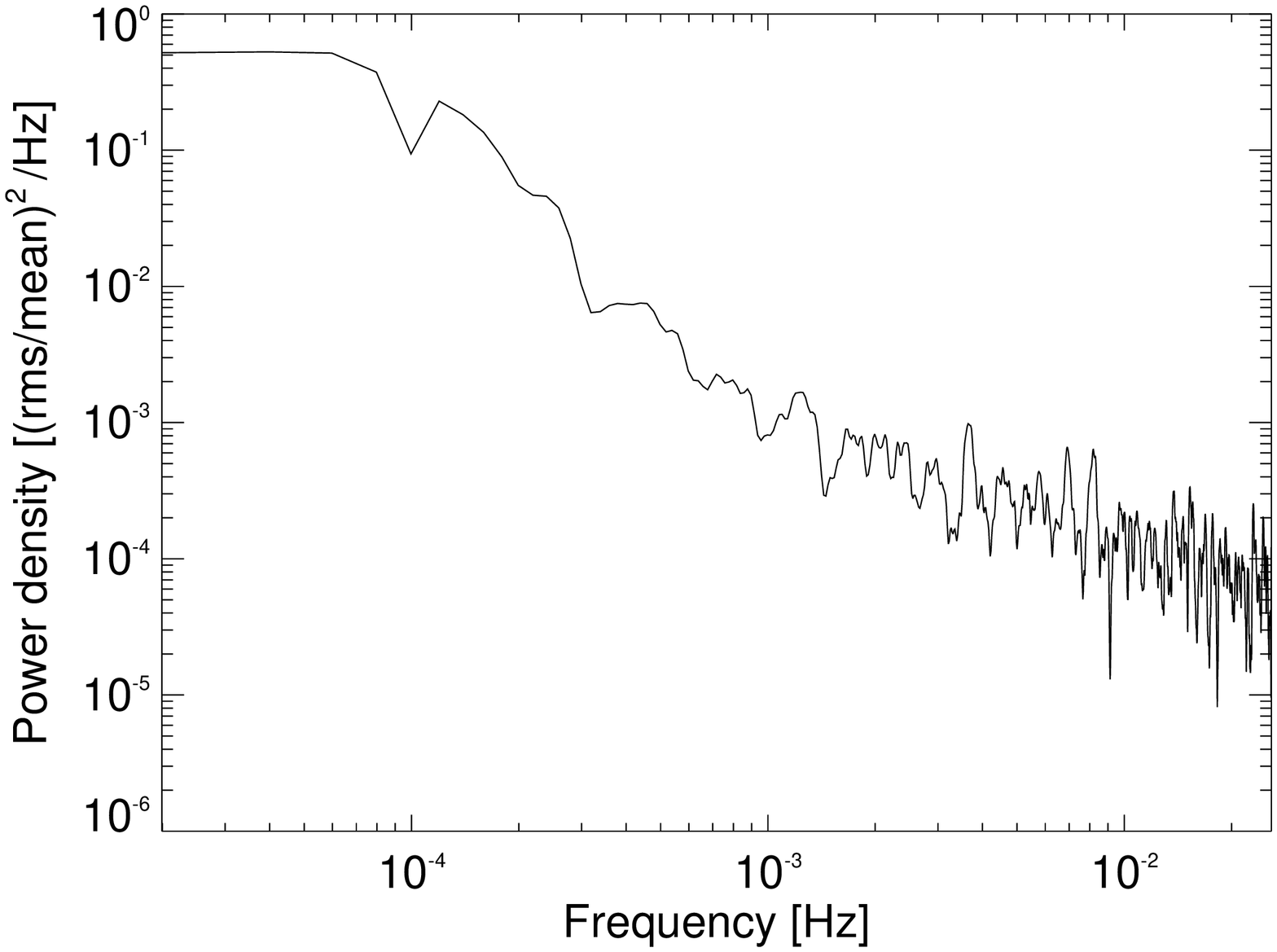} 
	\caption{Power spectra of the light curve of Mira in the first night of observations in $g'$ (left) and $u'$ (right).}
	\label{power_mira2}
	\includegraphics[trim=0cm 2cm 0cm 1cm, clip=true, width=0.3\textwidth, angle=-270]{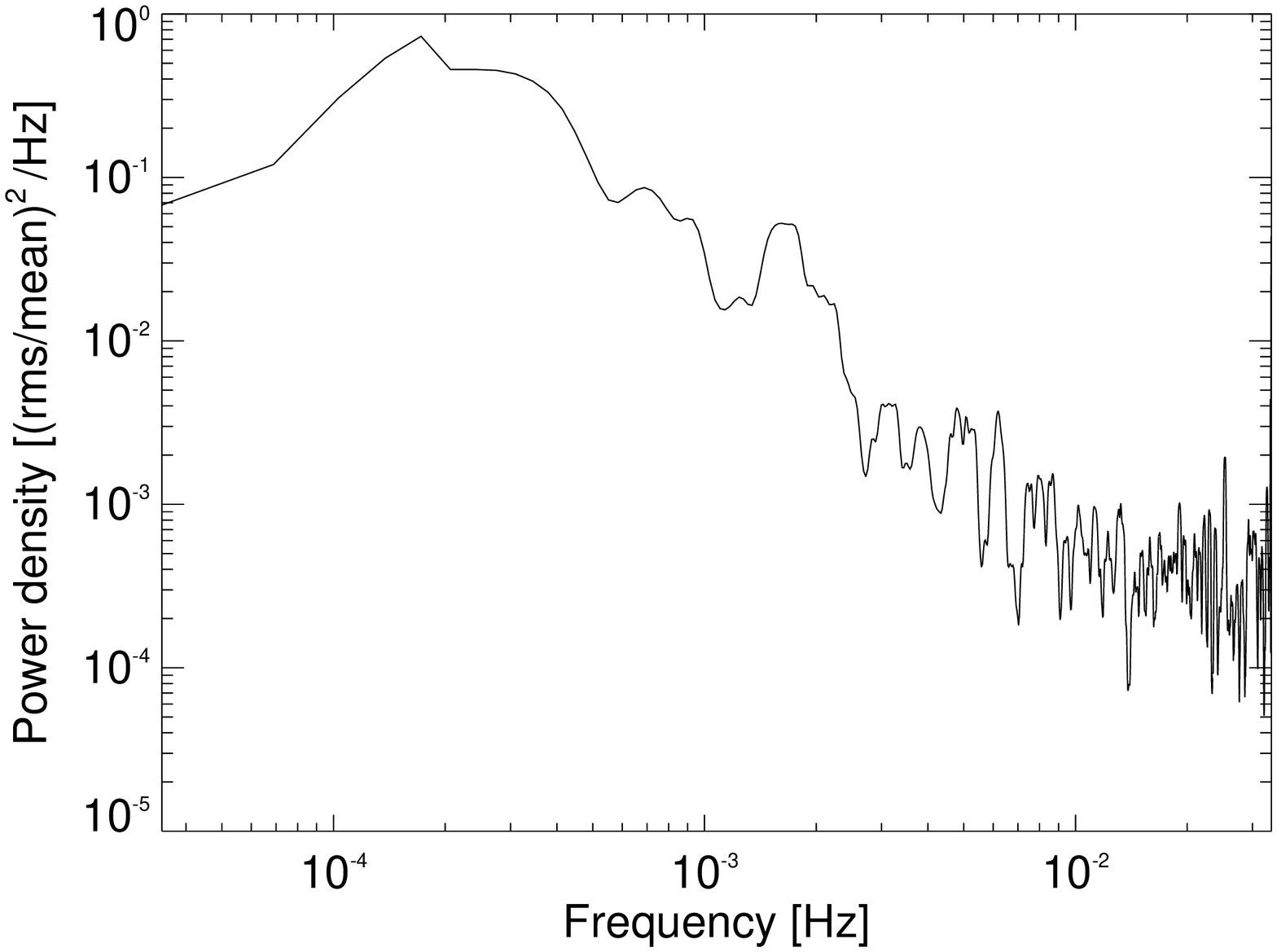}\qquad\includegraphics[trim=0cm 2cm 0cm 1cm, clip=true, width=0.3\textwidth, angle=-270]{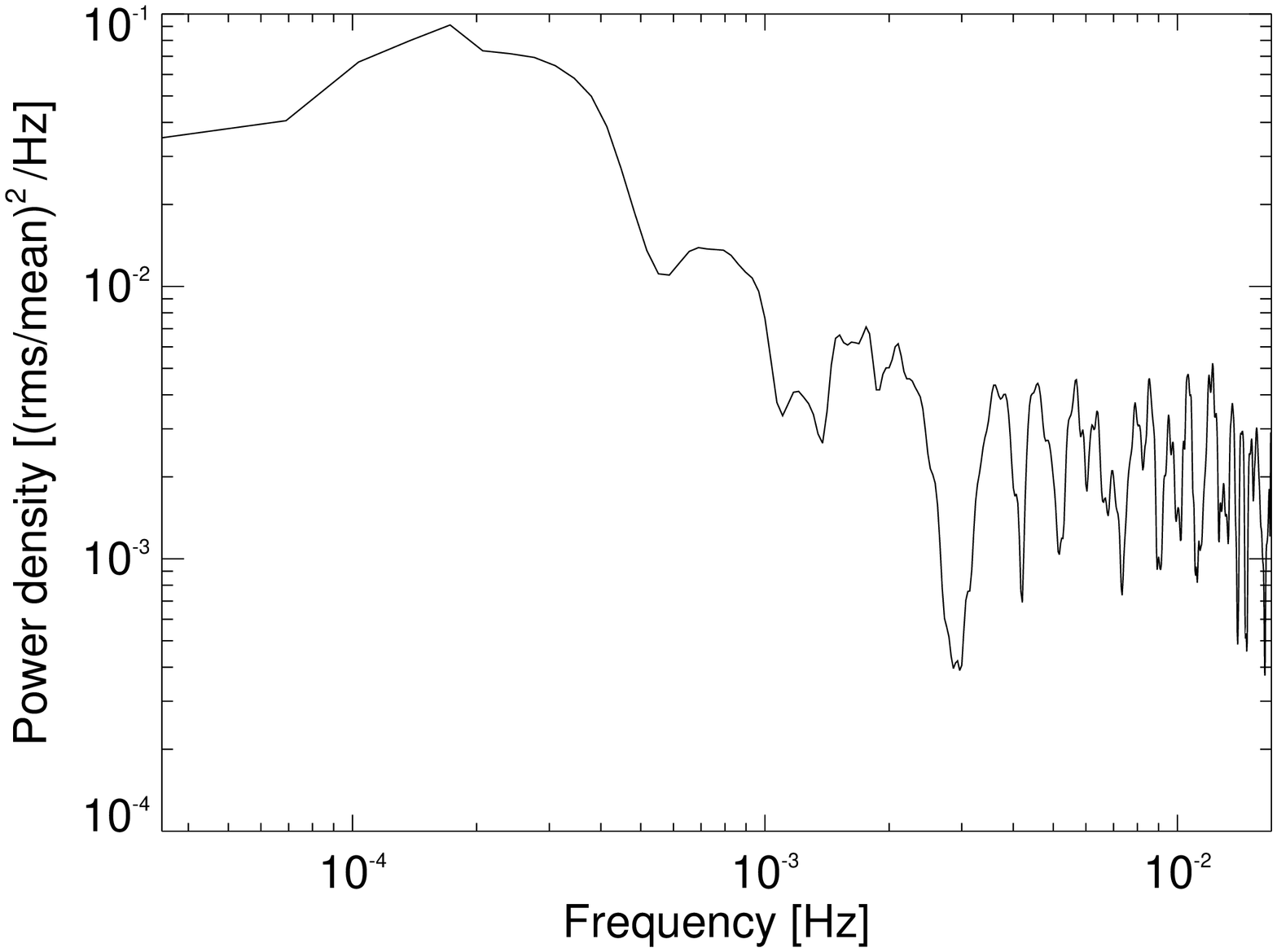} 
	\caption{Power spectra of the light curve of Mira in the second night of observations in $g'$ (left) and $u'$ (right).}
	\label{power_mira24}
\end{figure*}	

\subsubsection{Break Frequency}
Recently, \cite{Scaringi+15} have shown that the accretion physics of supermassive black holes at centres of active galactic nuclei is applicable to other types of accreting systems, despite the dissimilar nature of the accretor in these systems. Their rms-flux relation, which shows a linear relationship between the flux and the amount of flickering in an accreting system, seems to hold for all accreting systems from young and compact stellar objects to supermassive black holes. The relation shows a break in the power spectrum of the accreting system, and the location of the breaking point depends on the type of the accretor. For CVs, they found that this break occurs at $ 10^{-3} $Hz. Eq. (1) in \cite{Scaringi+15} calculates the break frequency ($\nu_{b} $) as follows:
\begin{equation}\label{logvb}
\log \nu_{b} = A \log R + B \log M + C \log \dot{M} + D
\end{equation}
where $R$, $M$, and $\dot{M}$ are the characteristic radius, accretor mass, and accretion rate, respectively and all are measured in cgs units. The constants in the equation are $A$ = −2.07, $B$ = 0.043, $C$ = 0.95, and $D$ = −3.07. 
For Mira~B, we are considering two cases; either it is a WD or an MS star. \\
If Mira~B is a WD, then we have $M$ = 0.6M$_\odot$, $R$ = 0.02R$_\odot$(using the mass-radius relation by \cite{Nauenberg72} and assuming that Mira~B is completely degenerate). The mass accretion rate is taken to be $ 10^{-8} $\,M$_\odot$ yr\textsuperscript{-1}. Thus, we calculate log ($\nu_{b} $) = $-3.7$, which is close to the calculated break frequency of CVs in \cite{Scaringi+15}. Both Fig. (\ref{power_mira2}) and (\ref{power_mira24}) show that the calculated break frequency is actually covered by our observations but no break could be seen in the power spectra at this frequency. \\
For the second case, where we assuming that Mira~B is an MS, we adopt the same values for both $M$ and $\dot{M}$ and for all constants. Using the mass-radius relation for MS stars \citep{DemircanKahraman91}, we calculate the corresponding radius R = 0.65R$_\odot$. The derived log ($\nu_{b}$) = $-6.8$, which is beyond the frequency coverage of our observations. \\
So, based on this relation we cannot distinguish between a WD and an MS companion.

\subsubsection{SED Fitting} \label{section:SED}
\begin{figure}
\centering
\includegraphics[width=1.0\linewidth]{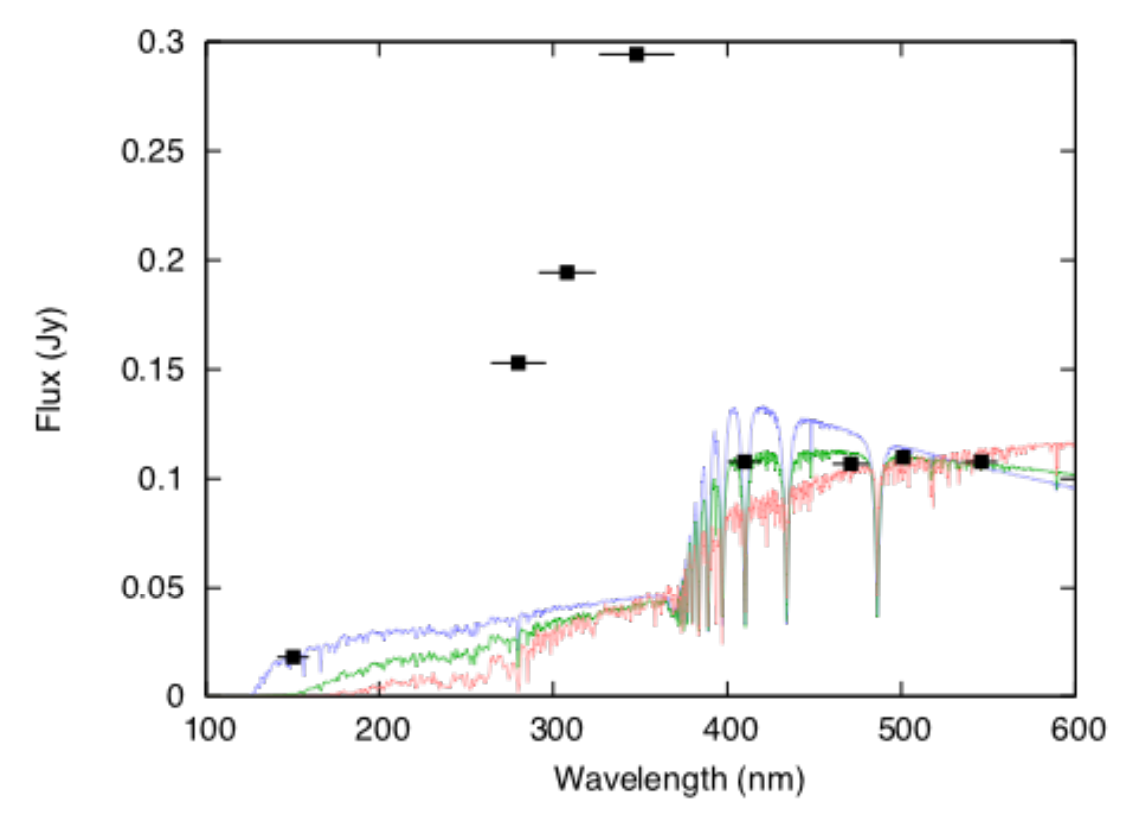}
\caption{Near\textendash UV spectrum of Mira~B: Black points from \protect\cite{Karovska+97}, solid lines are for BT-Settl models (e.g \citealt{Allard+13}) with [Z/H]\,=$-$0.5 dex, [$\alpha$/Fe]\,=+0.2 dex, log($g$) =+5.0 dex and $T_{e} = 10\,000$K (Blue), $8600$K (green) $7000$K (red), respectively.}	
\label{mira-b} 
\end{figure}

\indent The observed $U-B$ colour of Mira~B is $-1.0$, from the observations of \cite{Karovska+97}. For a stellar atmosphere, this colour implies a stellar temperature of 30\,000K. A star with this temperature would have a bolometric correction of $-3.16$. The measured $m_{V}=11.36$ and the Hipparcos distance of 92 pc would give a luminosity $L_{B}=3.5\rm L_\odot$. These temperature and luminosity are unlikely in the view of the lack of ionization and, in fact, are excluded by the SED.\\ 
\indent Fig. (\ref{mira-b}) shows the photometric observations of Mira~B from \cite{Karovska+97}. The ultraviolet photometry is well above the optical data and the $U-V$ colour is clearly not a good representative of the star. This behaviour indicates an accretion disk, where the Balmer jump is in emission (e.g. \citealt{Bapista+00}). We fitted the photometry using the BT-Settl models (Allard et al. 2013), excluding the three excess points, and found that the best fit is produced from a star with $T=8600$\,K and $L= 0.2 \rm L_\odot$. The luminosity in the Balmer continuum excess is 0.3L$_\odot$. The highest possible temperature is set by the bluest point ($\sim$ 150 nm) at 10,000K, and the lowest possible temperature is derived by excluding the B-band which could be affected by Balmer emission lines, and this gives a temperature of 7000\,K.\\
The range of temperatures indicates a star somewhat hotter than the Sun. If it was an MS star, it would therefore be more luminous than the Sun. This is inconsistent with the luminosity calculated above, and we can exclude such an MS star. This strongly favours a WD classification. The alternative is that all the photometric points are from the accretion disk, in which case the underlying star must be much less luminous, an M star or less.

\subsubsection{Accretion Rate and Luminosity}
\begin{figure}
	\begin{center}
		\includegraphics[width=1.0\linewidth]{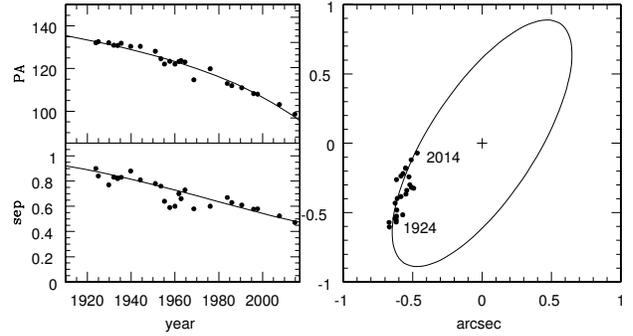}
		\caption{The orbit of Mira~B. The fit shows a circular orbit inclined at 67 degrees, with period 945\,yr.} \label{mirapos}
	\end{center}
\end{figure}

\indent The total accretion luminosity, approximately 0.5L$_\odot$, depends on the type of the companion. \cite{Soker04} gives the accretion rate as: \\

\begin{equation}\label{macc}\\
\begin{split}
\dot M_{acc} = 3\times10^{-7}\left( \frac{M_{WD}}{0.6\,\mathrm{M}_\odot }\right)^2 \left( \frac{v_{s}}{10\, \mathrm{ km\, s\textsuperscript{-1}}} \right)^{-4} \\
\left( \frac{a}{100\,\mathrm{AU}} \right)^{-2} \left( \frac{\dot M_{\ast} }{10^{-4}\,\mathrm{M}_\odot\, \mathrm{yr}^{-1} } \right) 
\end{split}
\end{equation}\\

\noindent where \\
$M_{WD}$ : mass of the accreting WD, we assume a typical value of 0.6M$_\odot$.\\
$v_{s}$: wind speed, we used typical speed of 10 km/s.\\
$\dot M_{\ast}$ : mass loss rate from Mira A, which is about $3\times10^{-7}$M$_\odot\, \mathrm{yr}^{-1}$.\\
$a$ : physical binary separation in AU.\\
In order to estimate the value of $a$ for Mira system, we used measured angular separation ($\theta$) between Mira~A and B. This conversion requires knowledge of the orbit. Mira~B has not been followed for long enough to establish an orbit but sufficient data is available for an initial fit. Published astrometric data are listed in Table \ref{pos_table}, supplemented with two new measurements obtained from \textit{HST} archival images. The data is plotted in Fig. (\ref{mirapos}).\\
We assume a circular orbit since the section of the orbit for which we have measurements do not show strong evidence for non-circular motion. However, an elliptical orbit is not ruled out. A reasonable fit is obtained for inclination $i=67^{\circ}$, polar direction at $-47^{\circ}$, $\theta=1.03$ arcsec, and a period of $P=945$ year. The polar direction and inclination can vary by few degrees. $\theta$ can be up to 1.2 arcsec if the inclination is taken as 72$^{\circ}$, or as low as 0.9 arcsec for an inclination of 64$^{\circ}$. In the first case, we find a period of 970 years, and in the latter a period of 860 days. At the extremes, the fit is notably worse for the earliest data.\\
Based on the value of 1.03 arcsec and a distance of 92\,pc, we find a physical separation between Mira~A and B of $ a=95$AU. 
The period and separation provides a combined mass for Mira AB which is in the range 0.75--1.5M$_\odot$. Larger values correspond to larger separations. If Mira~B is a WD, it is likely to have a mass larger than 0.5M$_\odot$. Mira~A is expected to have a mass larger than 0.6M$_\odot$, based on the fact that it has not yet
become a WD. This would favour a slightly larger value for the separation $a$ than chosen here, corresponding to a larger combined masses. However, the uncertainty introduced by the choice of a circular orbit is probably more significant. 

Using the above value of $a$ in Eq. (\ref{macc}), we find the predicted accretion rate to be $\dot M_{acc} = 9.97\times10^{-10}\mathrm{M}_\odot\, \mathrm{yr}^{-1}$. This Eq. assumes a Bondi-Hoyle-Lyttleton (BHL) accretion rate, which was found to give a lower accretion rates compared to models (\citealt{MP12}; \citealt{Huarte-Espinosa2013}). However, both models, showed that the accretion rates decrease with the increase in binary separation. In our calculations, we ignore the difference between BHL accretion rate and the models since we use a binary separation (95 AU) that is larger than binary separations used in these models.\\
The above estimated accretion rate is for a WD companion. A low-mass MS companion would give a slightly lower value because of its lower mass. \\
The correspondent accreting luminosity ($L_{acc}$) is calculated from the following equation \citep{Soker04}:
\begin{equation}\label{lacc}
L_{acc} = \frac{G\, \dot M_{acc}\, M_{WD} }{R_{WD}}
\end{equation}
where $R_{WD}$ is the radius of the accreting star. For a WD, we find $L_{\rm acc} \approx 0.94\,\mathrm L_\odot$. This is slightly higher than the observed accreting luminosity (0.5$\mathrm L_\odot$).
If the companion is an MS, then the accretion luminosity would be much less (about a factor of 100 less), which is much less than the observed luminosity. Moreover, the UV excess of 0.3$\mathrm L_\odot$ cannot be explained by an accretion around an MS. In combination with the SED temp of 7000-10\,000K, which is too high for a low mass MS star, we find the possibility of an MS companion very improbable.

\begin{table}
	\begin{center}
		\caption{Mira~B positions relative to Mira~A. Data is from \protect\cite{Baize80}, \protect\citet{Karovska+93,Karovska+97} and covers the 1983-1995 data values. The two new measurements are obtained from archival \textit{HST} data, and an ALMA measurement from \protect\cite{Vlemmings2015}, respectively.}
		\label{pos_table}
		\begin{tabular}{lcc}
			\hline
			\hline year & position angle ($^{\circ}$) & separation (arcsec)\\
			\hline 
			1923.88 &  132.1 & 0.90 \\
			1924.97 &  132.6 & 0.84 \\
			1929.73 &  132.1 & 0.77 \\
			1932.07 &  130.9 & 0.83 \\
			1933.77 &  130.8 & 0.82 \\
			1935.40 &  131.8 & 0.83 \\
			1939.73 &  130.4 & 0.88 \\
			1944.15 &  130.4 & 0.81 \\
			1951.01 &  128.1 & 0.78 \\
			1953.54 &  124.6 & 0.76 \\
			1955.22 &  122.1 & 0.64 \\
			1957.60 &  123.4 & 0.59 \\
			1959.96 &  122.1 & 0.60 \\
			1961.75 &  123.3 & 0.70 \\
			1962.76 &  123.7 & 0.66 \\
			1964.56 &  123.1 & 0.73 \\
			1968.61 &  114.7 & 0.58 \\
			1976.11 &  119.9 & 0.60 \\
			1983.88 &  113   & 0.67 \\ 
			1985.94 &  112   & 0.63 \\
			1990.52 &  111   & 0.61 \\
			1995.92 &  108.3 & 0.578 \\
			1997.75 &  108   & 0.58   \\
			2007.73 &  103.2 & 0.524 \\
			2014.82 &  98.6  & 0.472 \\
			\hline
			\hline
		\end{tabular}
	\end{center}
\end{table} 

\begin{figure*}
\includegraphics[trim=0cm 2cm 0cm 1cm, clip=true, width=0.3\textwidth, angle=-270]{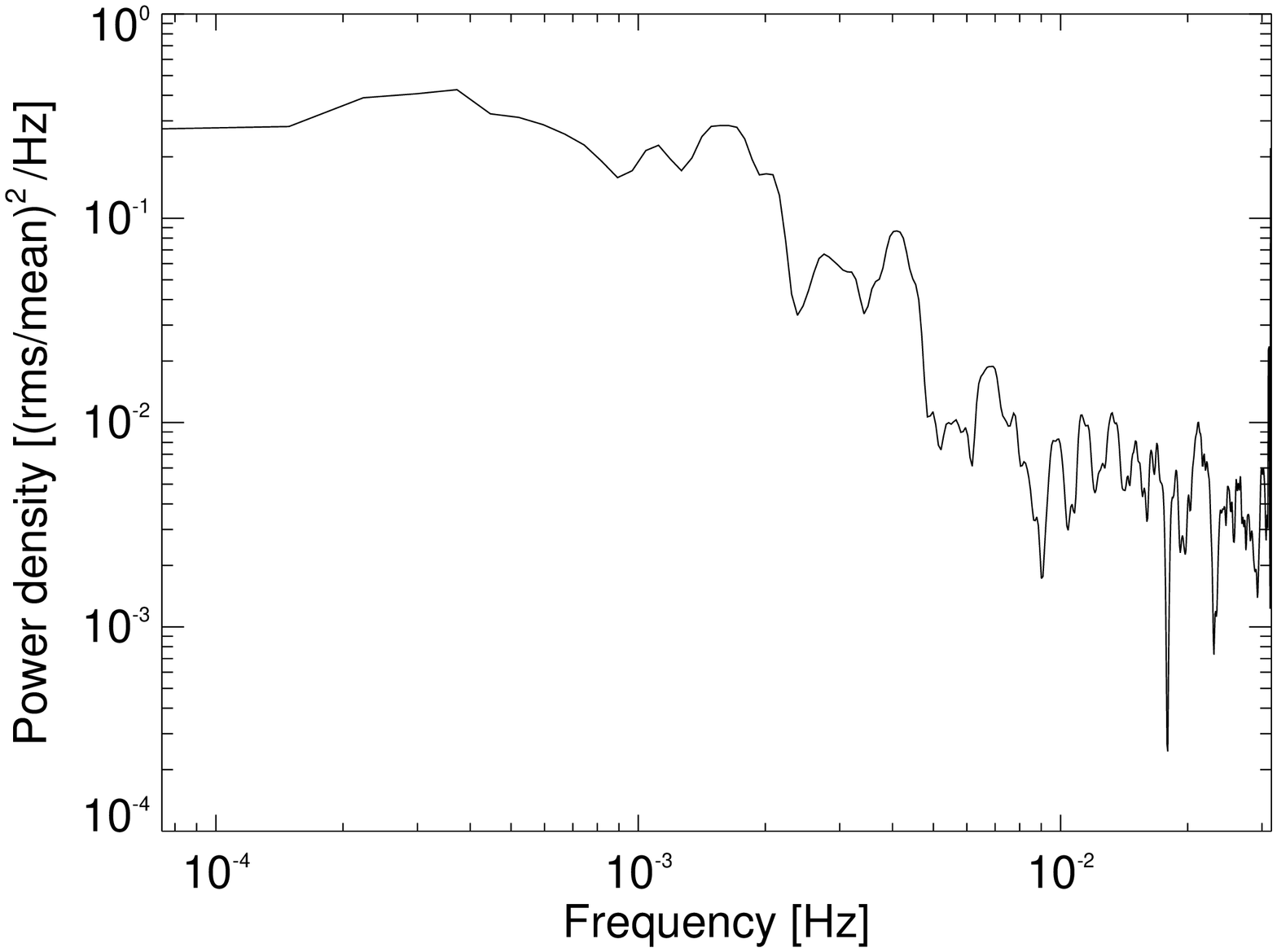}\qquad\includegraphics[trim=0cm 2cm 0cm 1cm, clip=true, width=0.3\textwidth, angle=-270]{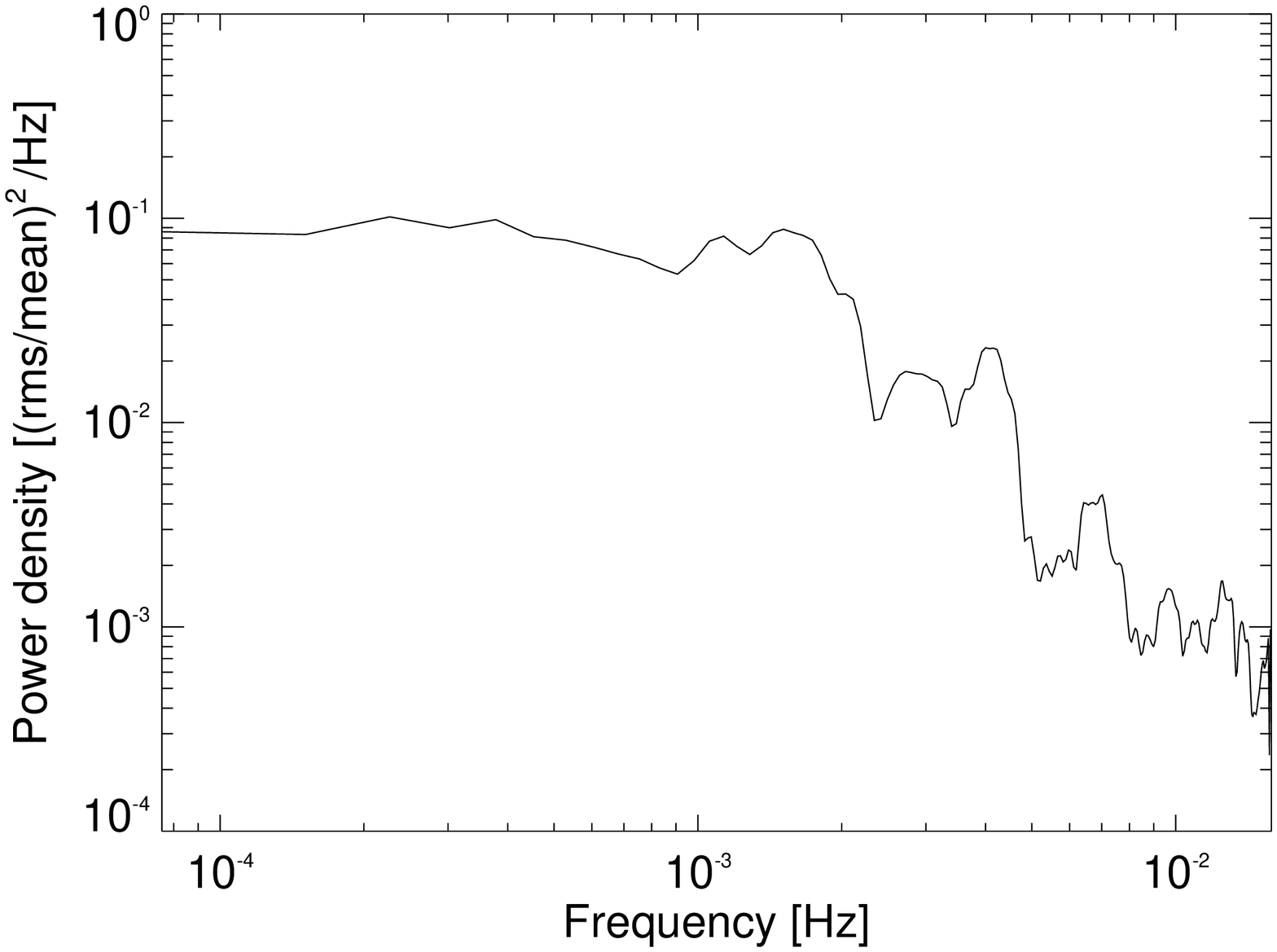} 
\caption{Power spectra of the light curve of Y Gem in $g'$ (left) and $u'$ (right). The comparison star is TYC 1369-542-1.}
\label{power_ygem-TYC5}
\end{figure*}	

\begin{figure*}
	\includegraphics[trim=0cm 2cm 0cm 1cm, clip=true, width=0.3\textwidth, angle=-270]{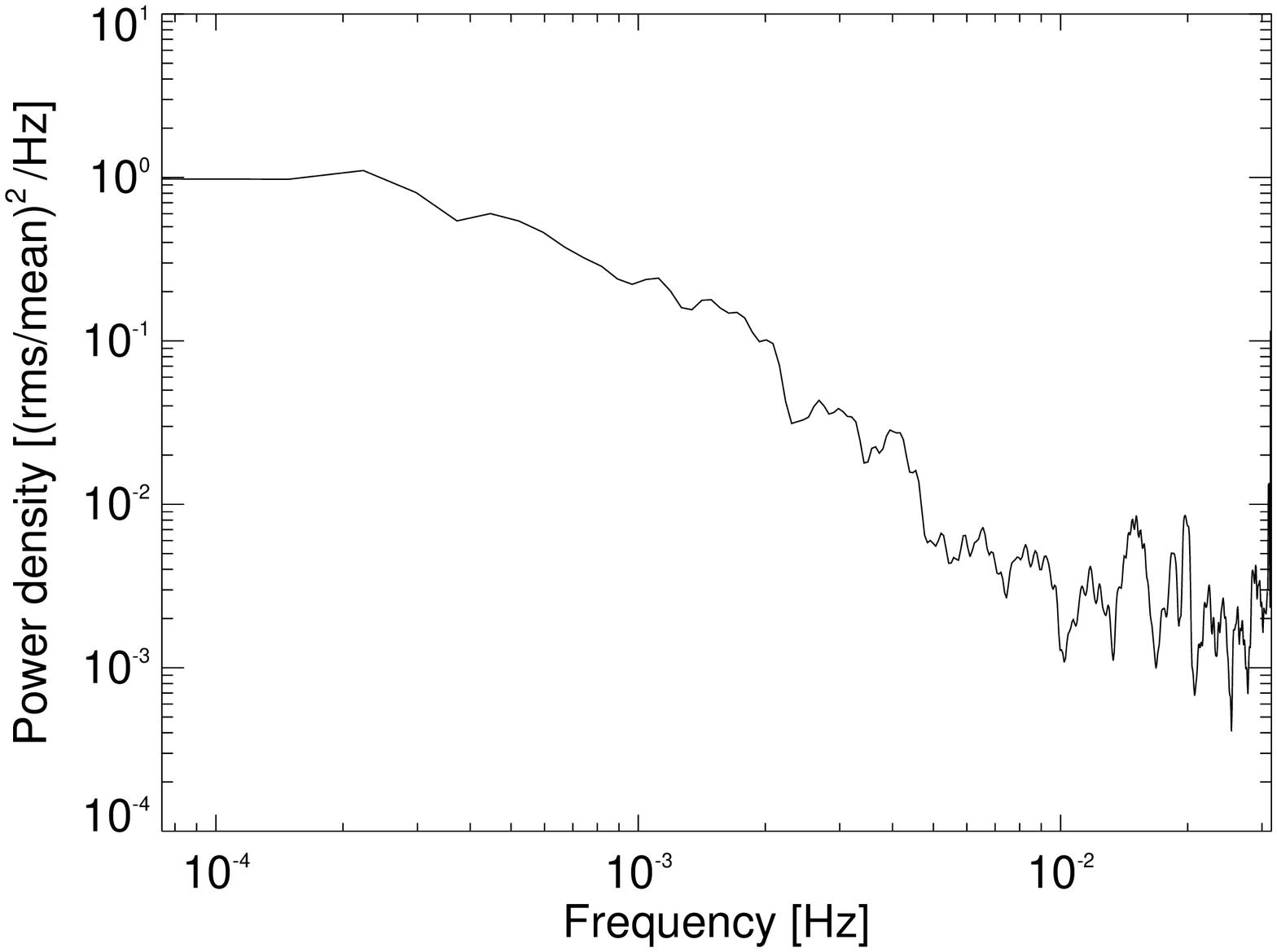}\qquad\includegraphics[trim=0cm 2cm 0cm 1cm, clip=true, width=0.3\textwidth, angle=-270]{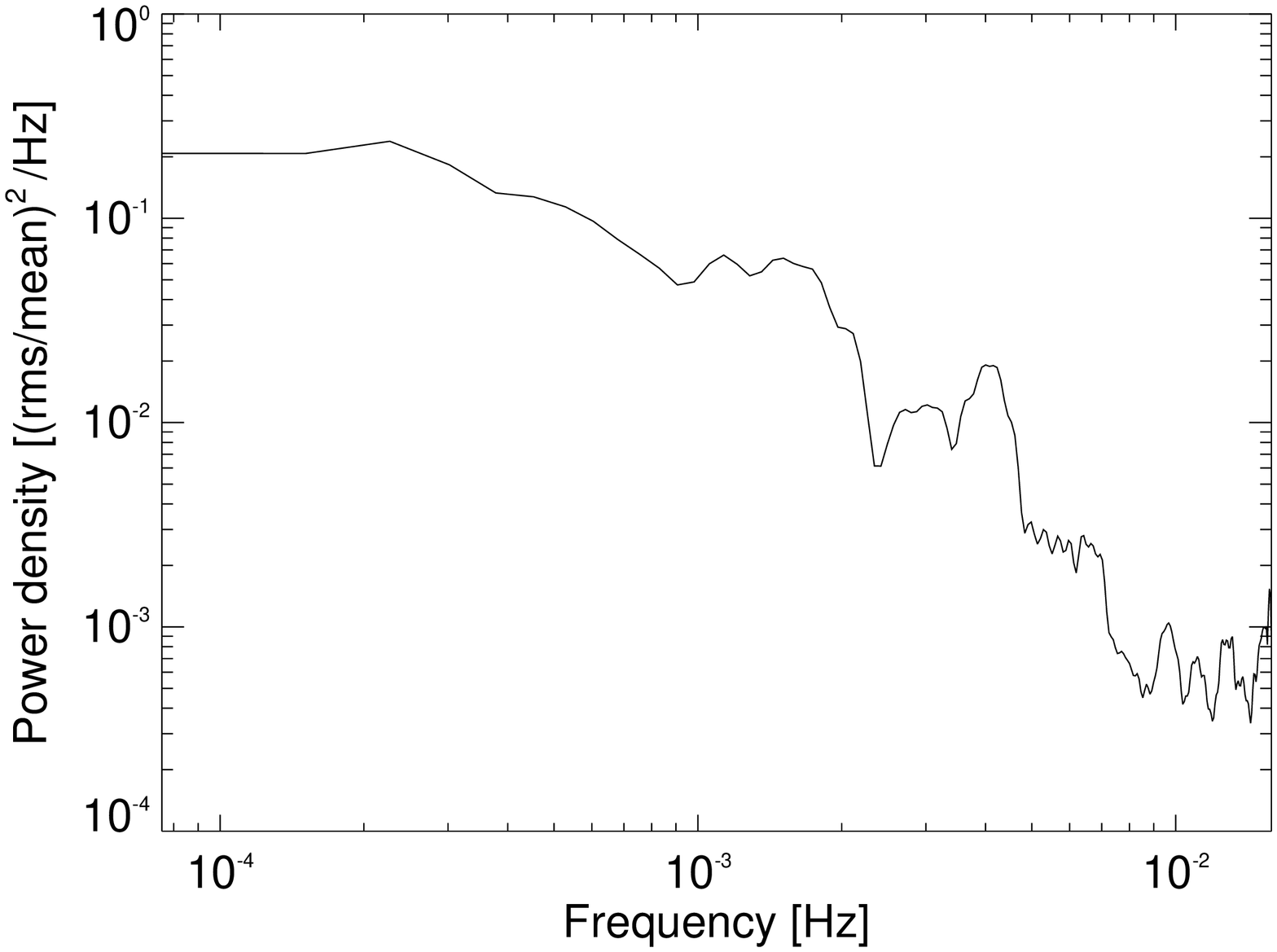} 
	\caption{Power spectra of the light curve of Y Gem in $g'$ (left) and $u'$ (right). The comparison star is TYC 1369-678-1}
	\label{power_ygem-TYC6}
\end{figure*}	


\section{Comparisons}
Flickering requires accretion at a high enough rate, and a WD companion. An MS companion would generate far less accretion luminosity. We observe flickering in two objects, Mira and Y Gem, which both have hot accretion disks in the UV spectra. On the other hand, we didn't observe flickering in both R~Aqr and V1016 Cyg. \\
There is no published value for the mass loss rate of Y Gem. However, it has an average Infrared excess of about 1.35 \citep{McDonald+12}, which suggests a quite low mass loss rate ($\sim10^{-7} \rm M_\odot\,yr^{-1} $ or less). The accretion luminosity of 2.4$\rm L_\odot$ calculated by \cite{Sahai2011} is actually higher than what we calculate for Mira, however it is probably quite uncertain because the distance has a high uncertainty. If this accretion luminosity is correct and Y Gem has a very low mass loss rate, this would suggest that the orbital separation is smaller than it is for Mira.\\
\indent Eq. (\ref{macc}) shows that the accretion rate is independent of the type (i.e. radius) of companion, as long as the mass is the same. However, Eq. (\ref{lacc}) shows that the accretion luminosity is a strong function of companion type. A high accretion luminosity per unit mass also heats the accretion disk and increases the contrast with the Mira primary. Although flickering was, thus, mainly expected for WD companions, Y Gem has an accretion luminosity within range of what is achievable from an MS companion.  \\
\indent R~Aqr has a higher mass loss rate than Mira ($5 \times 10^{-6}\,\rm M_\odot\,yr^{-1}$; \citealt{Bujarrabal+10}) and has a companion at 200\,AU which drives the jet outflows. The absence of flickering in R~Aqr may suggest that its companion is an MS star, rather than a WD as suggested by Nichols et al. (2007): the accretion rate is expected to be similar to that of Mira, as the stellar mass loss rate is an order of magnitude higher and the companion a factor of 3 more distant. However, another possibility is that the accretion disk is obscured by the Mira wind. \\
\indent V1016 Cyg has a WD companion as shown by its nova outbursts. It has a high mass-loss rate of $\dot M = 2 \times 10^{-6} \rm M_\odot\, \mathrm{yr}^{-1}$ \citep{SS15} and its Mira period is 465 days. The orbital parameters for V1016 Cyg are not known, so its accretion rate cannot be calculated. However, it is likely higher than that in Mira due to the high mass loss rate and the recent slow nova outburst. Its companion is certain to be a WD. Flickering would be expected, and the fact that it is not seen suggests there are factors which suppress flickering for these parameters. These factors could also apply to R~Aqr. \\
\indent It is not clear why both R~Aqr and V1016 Cyg do not show flickering. However, we note that our observations of R~Aqr were short and taken under non-ideal circumstances. Repeating these observations might be helpful. It is notable that the two clear cases of flickering are for the two stars with the lowest mass loss rates. However, the sample is too small to discuss the significance of this finding. 

\begin{table}
\begin{minipage}[b]{0.5\textwidth}\centering
\caption{Properties of power spectra. Slopes and fractional RMS values (column 4 and 5, respectively) are calculated between $10^{-2} - 10^{-3}$Hz (See \S \ref{section:power.spectra}).}
\label{power spectra}
\centering
\setlength{\tabcolsep}{2.5pt}
\begin{tabular}{llccc}
\hline
\hline
Target			&	Comparison			&	Band	&	Slope  &	 Fr. RMS (\%) \\
\hline
V1016 Cyg	&	TYC 3141-577-1			&	$r'$		&	-1.99	&	2.03  	\\
													&	& $u'$			&	-0.39	& 0.26	\\
	&USNO-A1.0 1275-13058156		&$r'$			&-1.74		&	2.00\\
													&	&$u'$&		-0.73		&		0.38	\\
													
RAqr	&  									   &	$g'$		&-0.45		&	1.77	\\
Mira\textsuperscript{1}	&HD14411	&$g'$&-1.86	&0.19	\\
												&	&$u'$&	-0.85	&	0.18		\\
Mira\textsuperscript{2}	&HD14411&$g'$&-1.97	&0.70	\\
												&	&$u'$	&	-0.33	&	0.50	\\
YGem&TYC 1369-542-1					&$g'$&	-1.86	&2.10	\\
								&					&$u'$&	-1.95	&		1.13	\\
		&TYC 1369-678-1					&$g'$&-1.62&1.70 \\
							&					&	$u'$	&	-1.94	&	0.98	\\
\hline
\hline
\end{tabular}
\end{minipage}
\begin{tablenotes}\footnotesize
\item[*] (1) First night of observations. (2) Second night of observations. 
\end{tablenotes}
\end{table}

\section{Conclusion}
We have carried out observations of four AGB stars known to have binary companions, where the companion is likely accreting: Mira, Y Gem, R~Aqr and V1016 Cyg. We used ULTRACAM to observe our targets in order to investigate flickering at very short timescales (minutes or less). We found evidence for flickering in both Mira and Y Gem, but not in R~Aqr and V1016 Cyg. \\
\indent Mira was observed at two nights when it was 1 mag above its minimum and the typical amplitude of the fast fluctuations in $g'$ band is 0.005 mag and even less in $u'$. The power spectrum of Mira shows a good detection of fluctuations but no evidence for periodicity. \\ 
We found that the rms variations from Mira system between $10^{-3}$ to $10^{-2}$ Hz are 0.19\% in $g'$ and 0.18\% in $u'$ from the first night of observations, and 0.7\% and 0.5\% in $g'$ and $u'$ in the second night (Table \ref{power spectra}).\\
To calculate rms variations from Mira~B only, we assume that all flickering is coming from Mira~B and correct for Mira~A flux. We found that the rms on Mira~B becomes 2.4\% in $g'$ and 0.5\% in $u'$ in the first night, and 8.8\% and 1.1\% in $g'$ and $u'$ in the second night. The average value of rms variations in CVs over the same frequency range are 2.4\% in $g'$ and 4.5\% in $u'$ \citep{BarrosPhD}. Our calculations are smaller than this average in $u'$, and similar to the average in $g'$ for the first night only. The fractional rms variabilities of each night are quite comparable for both bands, although $u'$ is expected to give higher values. This is seen in light curves but is not mirrored in these calculations. \\
\indent We studied the nature of Mira~B using three different ways:
\begin{enumerate}
\item The rms-flux relation by \cite{Scaringi+15} shows that the bending frequency in power spectra of accreting systems is about $ 10^{-3} $Hz in CVs. We calculate a similar value, assuming that Mira~B is a WD. However, we could not confirm this conclusion since the break does not appear in any power spectra of Mira. At the same time, we find that our observations are not long enough to support a main-sequence nature as the calculated breaking point is not covered by our observations.
\item The Near-UV spectrum of Mira~B (\citealt{Karovska+97}) is found to be compatible with a star with temperature that ranges between 7000K and 10\,000K. In both cases Mira~B should be hotter than the Sun. It is unlikely to be an MS star because this high temperature requires Mira~B to be more luminous than the Sun, and our calculations found that the luminosity of Mira~B is only 0.5L$_\odot$. This might support the WD nature of Mira~B.
\item Using accretion luminosity equation for a WD \citep{Soker04}, we find that the calculated accretion luminosity is close to the observed. This strongly favours a WD companion. 
\end{enumerate}

\indent Y Gem was observed with two comparison stars in the sky. The differential photometry of Y Gem in $g'$ band shows large fluctuations with a timescale of 10 minutes and even larger in the $u'$ band. Short-term variations are more noticeable in $u'$, which suggest that the companion is bluer than Y Gem in ($u' - g'$). An estimation shows that the companion contributes only 4.4\%\ of the total flux in $g'$, which corresponds to a fractional rms of about 20\%. Compared to Mira, Y Gem has an equal or less mass loss rate and a higher accretion luminosity. This suggests that Y Gem has a smaller binary separation than Mira. However, this is not certain as the distance used in calculating the accretion luminosity has high uncertainty. \\
\indent R~Aqr shows no signs for flickering in our observations, which may indicate that its companion is an MS. The accretion rate is expected to be similar to Mira since R~Aqr has a higher mass loss rate and larger binary separation. If the companion is a WD, then flickering could be hidden by the stellar wind of the primary.\\
\indent We couldn't observe flickering in V1016 Cyg light curves, although its companion is confirmed to be a WD and the giant has a high mass-loss rate. The latter together with the recent slow nova eruption hint that the accretion rate of V1016 Cyg is likely higher than Mira. Unfortunately, the accretion rate couldn't be calculated since the orbital parameters are not known.

\section*{Acknowledgments}
We acknowledge the variable star observations from the AAVSO International Database contributed by observers worldwide and used in this work. We are grateful for constructive suggestions by the anonymous referee. \\

\bibliographystyle{mn2e}
\bibliography{ucam}

\clearpage
\appendix 
\begin{figure}
\section{Power Spectra for R Aqr}	
\begin{minipage}[h]{\textwidth}
	\centering
\includegraphics[trim=0cm 2cm 0cm 1cm, clip=true, width=0.4\textwidth, angle=-270]{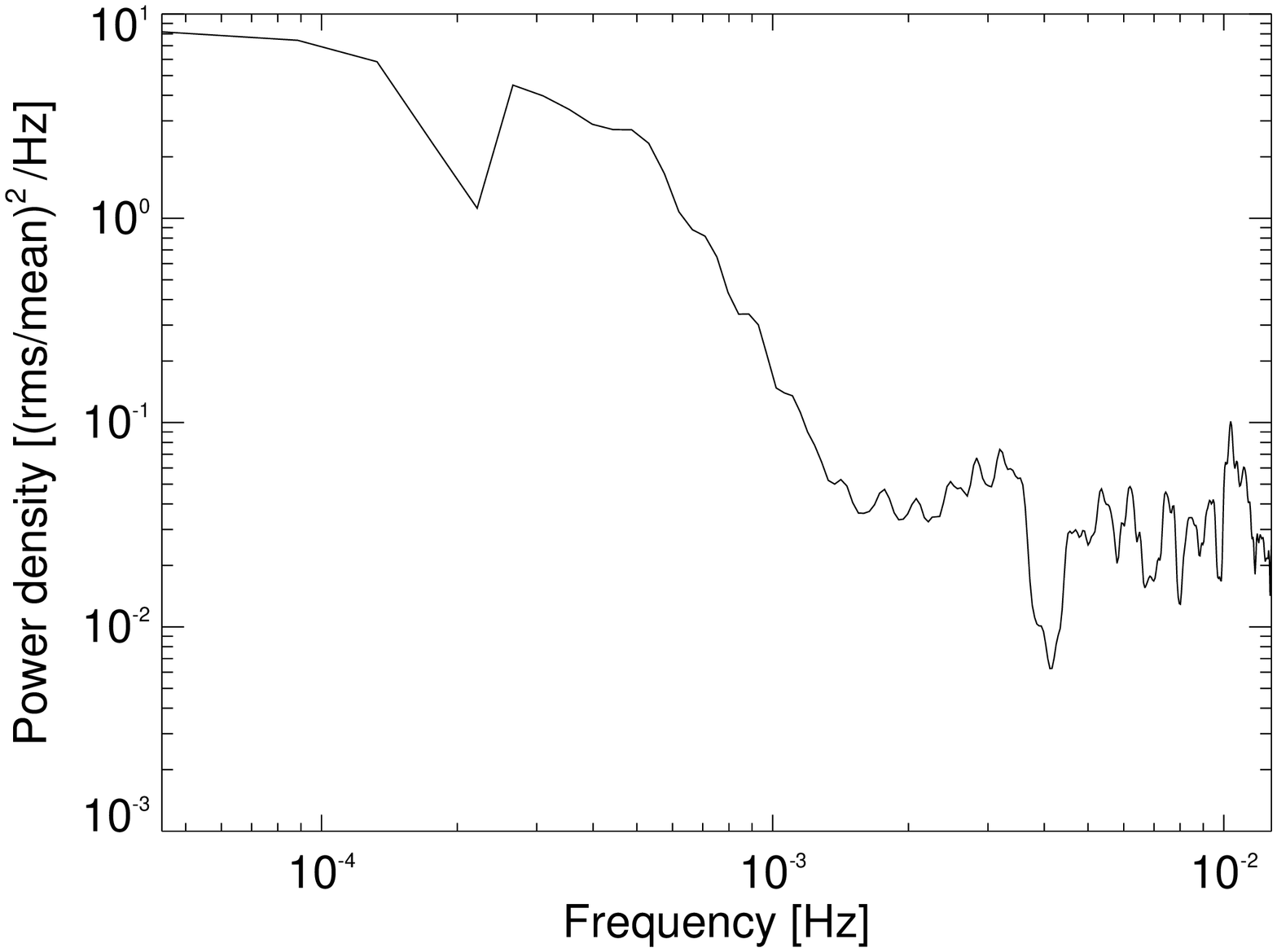} \\
\includegraphics[trim=0cm 2cm 0cm 1cm, clip=true, width=0.4\textwidth, angle=-270]{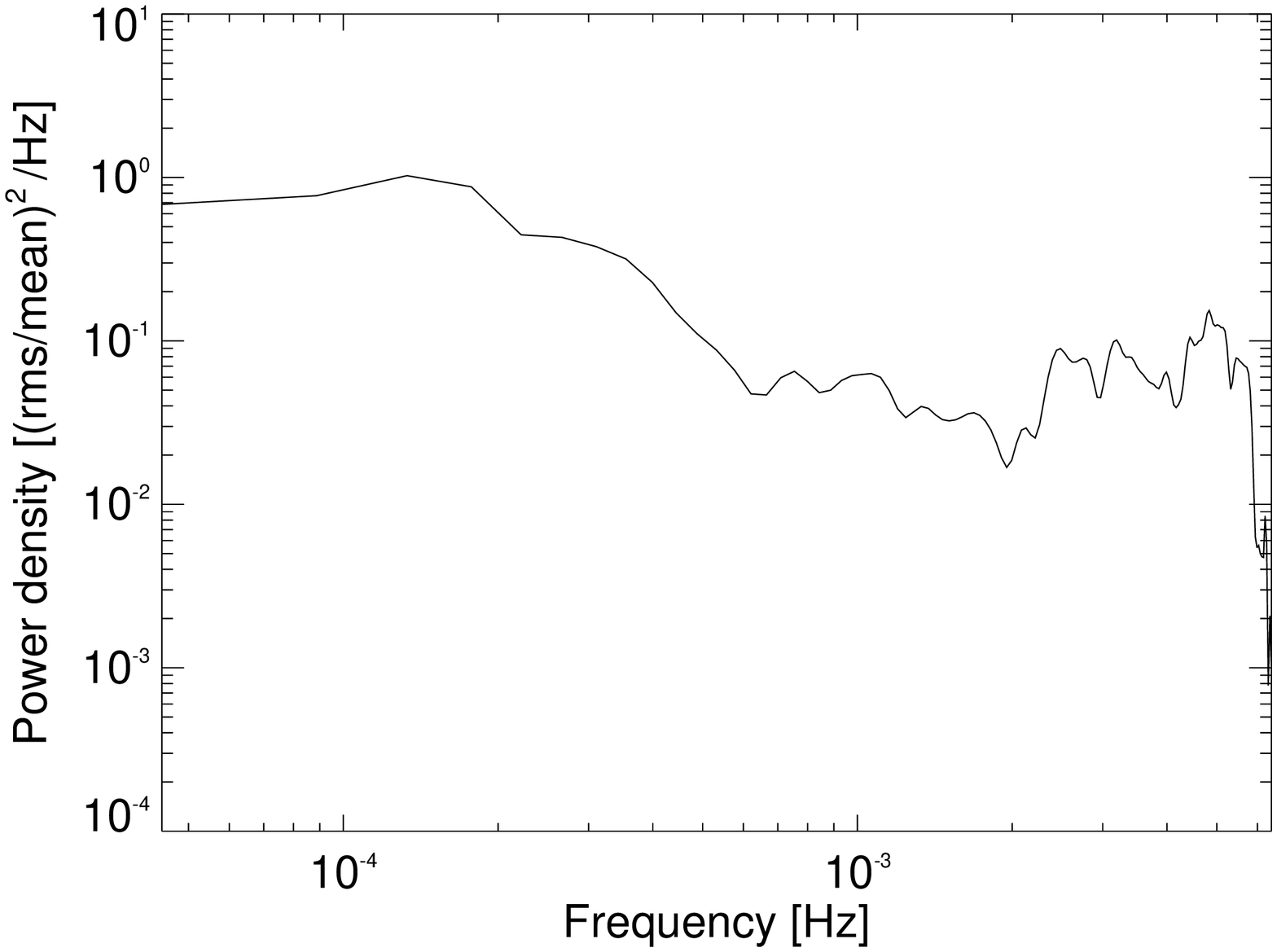}
\caption{Power spectra of the light curve of R~Aqr in $g'$ (top) and $u'$ (bottom).} 
\label{power_raqr_left}
\end{minipage}
\end{figure}

\clearpage
\newpage

\begin{figure}
\section{Power Spectra for V1016 Cyg}	
\begin{minipage}[c]{\textwidth}
\centering
\includegraphics[trim=0cm 2cm 0cm 1cm, clip=true, width=0.4\textwidth, angle=-270]{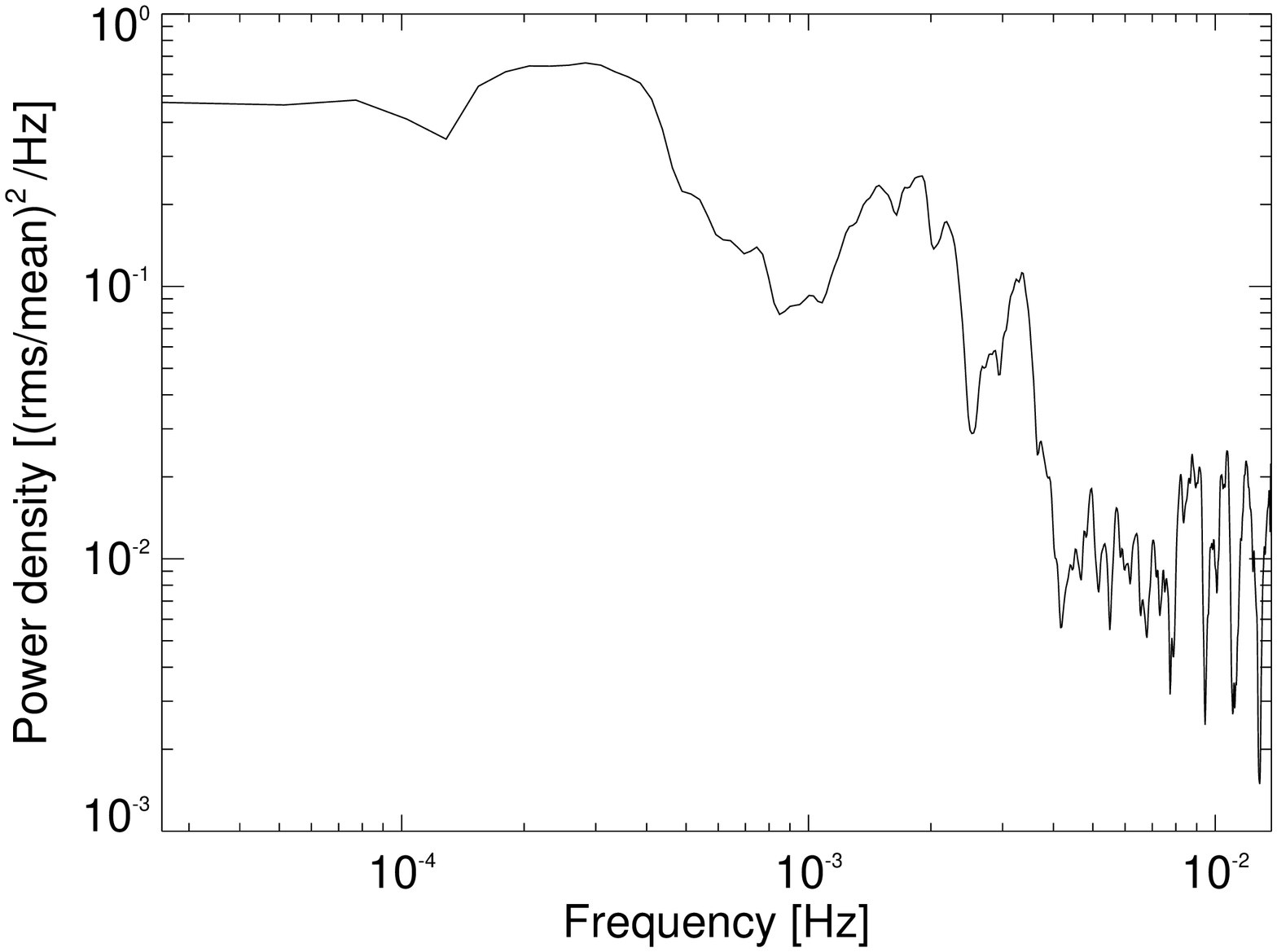}\\
\includegraphics[trim=0cm 2cm 0cm 1cm, clip=true, width=0.4\textwidth, angle=-270]{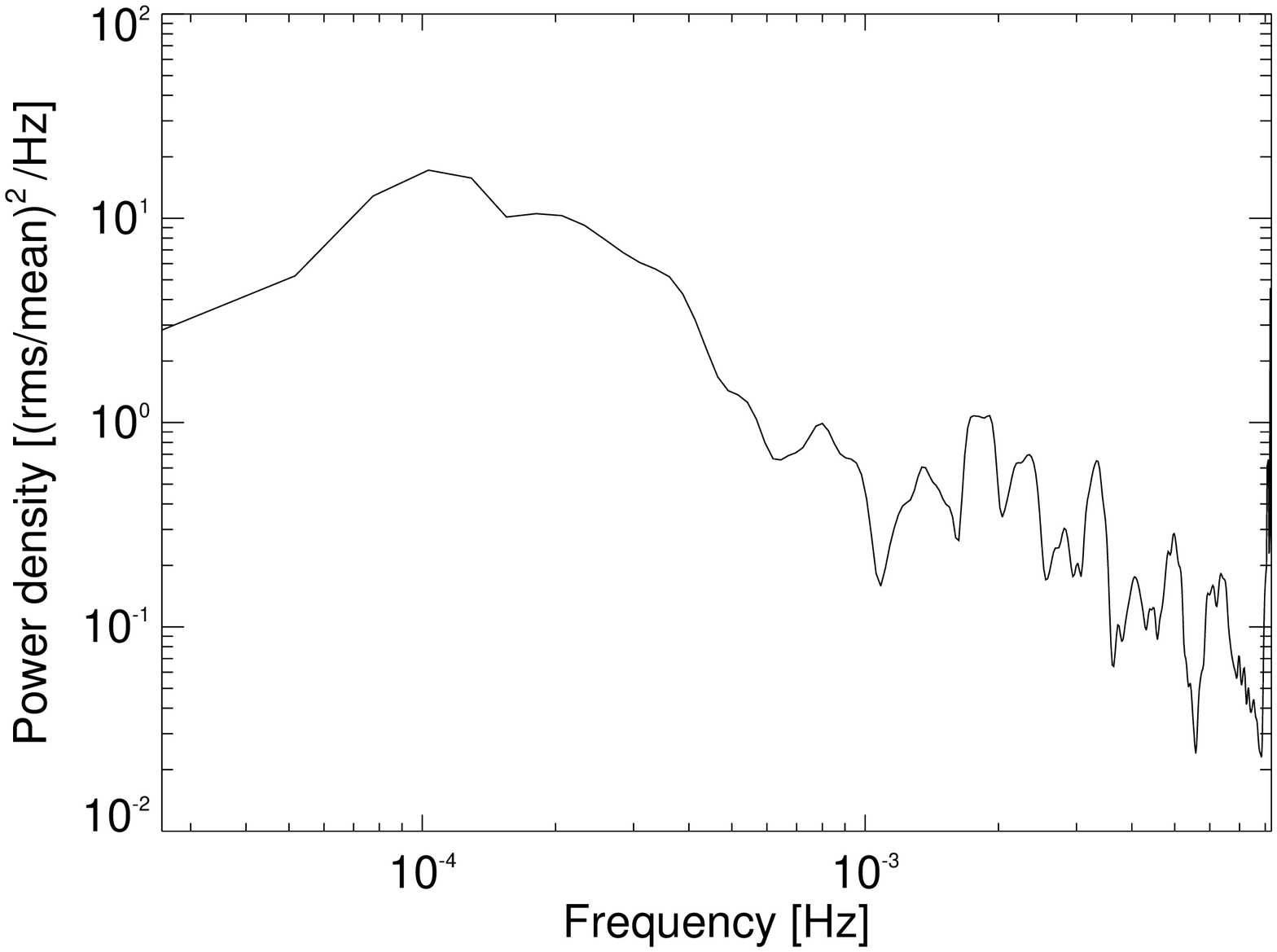}\\
\includegraphics[trim=0cm 2cm 0cm 1cm, clip=true, width=0.4\textwidth, angle=-270]{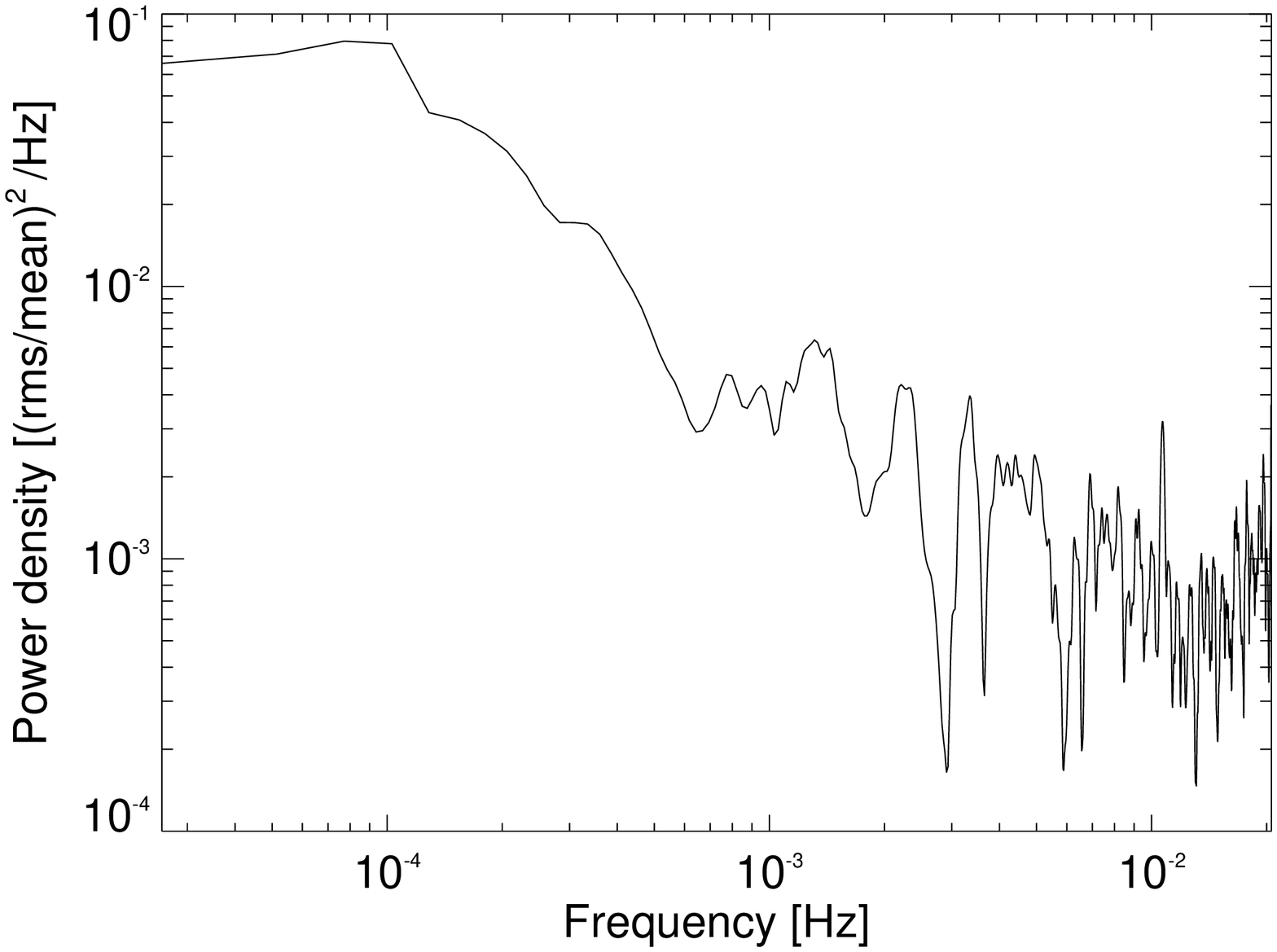}
\caption{Power spectra of the light curve of V1016 Cyg in $r'$, $g'$ and $u'$. The comparison star is USNO-A1-0-1275-13058156.} 
\label{Power_V1016Cyg1}
\end{minipage}
\end{figure}
\clearpage
\newpage		
\begin{figure}
\begin{minipage}[c]{\textwidth}
		\centering
\includegraphics[trim=0cm 2cm 0cm 1cm, clip=true, width=0.4\textwidth, angle=-270]{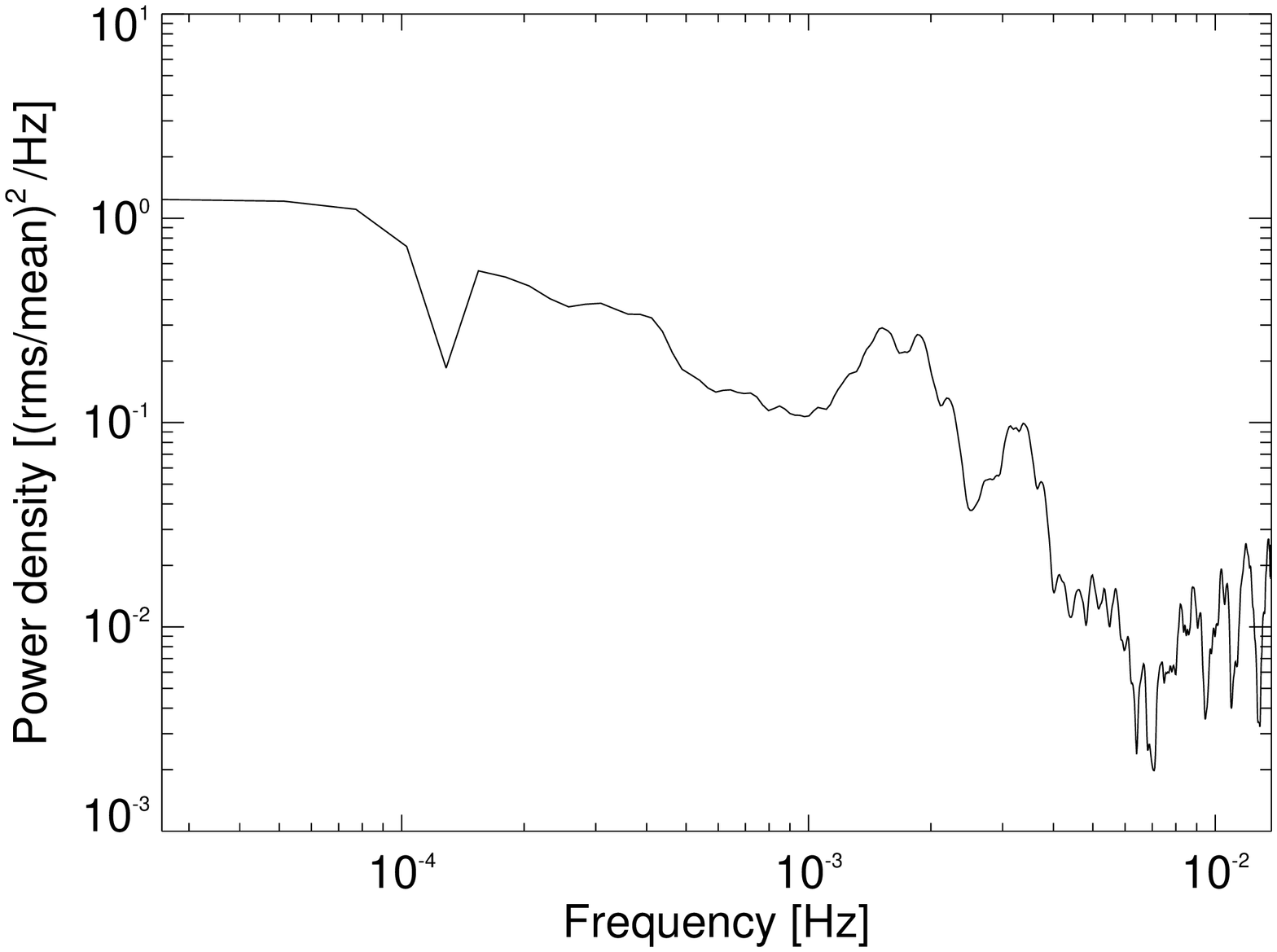}\\
\includegraphics[trim=0cm 2cm 0cm 1cm, clip=true, width=0.4\textwidth, angle=-270]{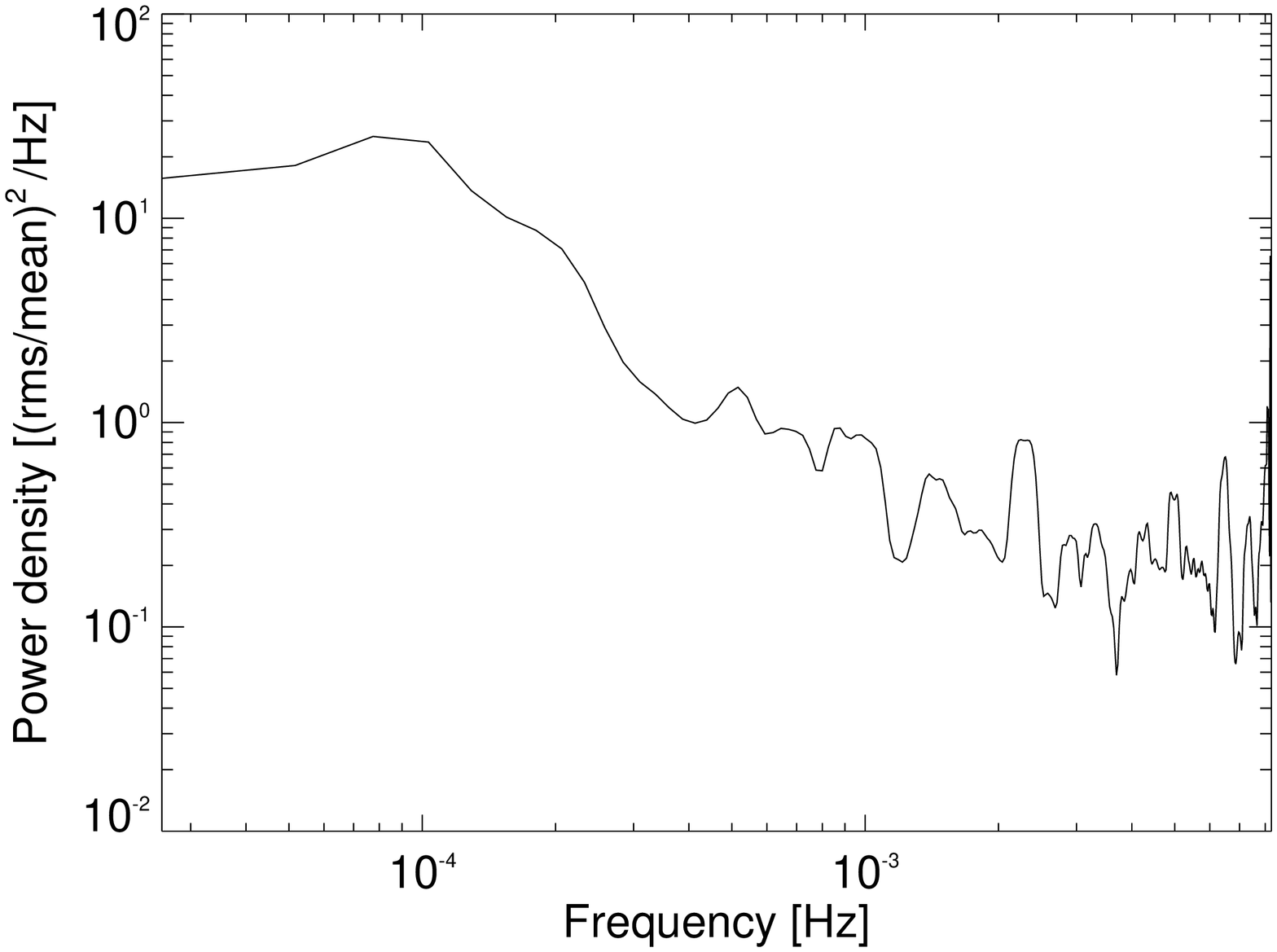}\\
\includegraphics[trim=0cm 2cm 0cm 1cm, clip=true, width=0.4\textwidth, angle=-270]{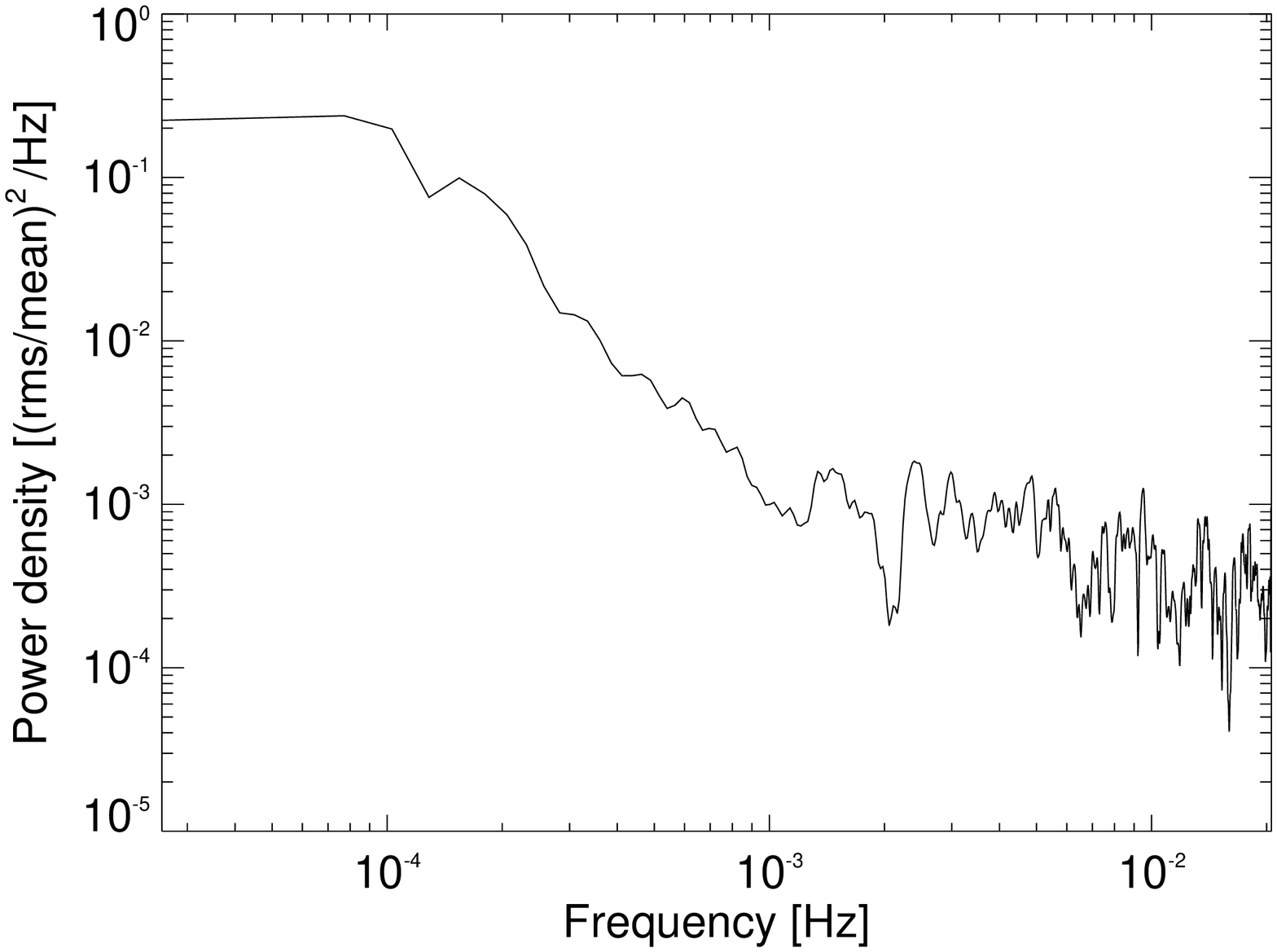}
\caption{Power spectra of the light curve of V1016 Cyg in $r'$, $g'$ and $u'$. The comparison star is TYC 3141-577-1.} 
\label{power_V1016Cygdm2}
\end{minipage}
\end{figure}	
\clearpage
\newpage		

\end{document}